\PassOptionsToPackage{table,xcdraw}{xcolor}
\documentclass[11pt,a4paper]{article}
\pdfoutput=1
\usepackage{jheppub}
\usepackage{gensymb}
\DeclareSymbolFont{matha}{OML}{txmi}{m}{it}
\DeclareMathSymbol{\varv}{\mathord}{matha}{118}
\usepackage{amsmath}
\usepackage{amssymb}
\usepackage{graphicx}
\usepackage{subfigure}
\usepackage{pstricks}
\usepackage{bm}
\usepackage{pbox}
\usepackage{placeins}
\usepackage{booktabs}
\usepackage[T1]{fontenc}
\usepackage{footnote}
\usepackage{pdfpages}
\usepackage{hhline}
\usepackage{multirow}
\usepackage{multicol}
\usepackage{enumitem}
\usepackage[toc,page]{appendix}
\allowdisplaybreaks
\usepackage{comment}
\usepackage{float}
\usepackage{slashed}
\usepackage{xcolor}
\usepackage{textcomp}
\usepackage{gensymb}
\usepackage{multirow}
\usepackage[numbers]{natbib}
\usepackage{notoccite}
\usepackage{tabu}
\usepackage{soul,xcolor}
\setstcolor{red}

\usepackage{rotating}
\usepackage{tabularx}
\usepackage[labelfont=bf]{caption} 

\FloatBarrier

\usepackage[many]{tcolorbox}

\renewcommand{\thetable}{\Roman{table}}

\definecolor{MyDarkBlue}{rgb}{0.1, 0.1, 0.8} 
\definecolor{MyLightBlue}{rgb}{0.22,0.51,0.9}
\definecolor{MyGreen}{rgb}{0.0, 0.5, 0.0}
\definecolor{BrickRed}{rgb}{0.8, 0.25, 0.33}

\RequirePackage{hyperref}
\hypersetup{colorlinks, citecolor=MyGreen,linkcolor=cyan, urlcolor=MyLightBlue}

\makeatletter
\gdef\@fpheader{}
\makeatother
\begin{document}

\title{\bf Quark-lepton Yukawa ratios and nucleon
decay in SU(5) GUTs with type-III seesaw 
}

\author[]{Stefan Antusch,}
\author[]{Kevin Hinze,}
\author[]{and Shaikh Saad}

\affiliation[]{Department of Physics, University of Basel, Klingelbergstrasse\ 82, CH-4056 Basel, Switzerland}

\emailAdd{stefan.antusch@unibas.ch, kevin.hinze@unibas.ch, shaikh.saad@unibas.ch}
\abstract{
We consider an extension of the Georgi-Glashow SU(5) GUT model by a 45-dimensional scalar and a 24-dimensional fermionic representation, where the latter leads to the generation of the observed light neutrino masses via a combination of a type I and a type III seesaw mechanism. Within this scenario, we investigate the viability of predictions for the ratios between the charged lepton and down-type quark Yukawa couplings, focusing on the second and third family. Such predictions can emerge when the relevant entries of the Yukawa matrices are generated from single joint GUT operators (i.e.\ under the condition of \textit{single operator dominance}). We show that three combinations are viable, (i) $y_\tau/y_b=3/2$, $y_\mu/y_s=9/2$, (ii) $y_\tau/y_b=2$, $y_\mu/y_s=9/2$, and (iii) $y_\tau/y_b=2$, $y_\mu/y_s=6$. We extend these possibilities to three toy models, accounting also for the first family masses, and calculate their predictions for various nucleon decay rates. We also analyse how the requirement of gauge coupling unification constrains the masses of potentially light relic states testable at colliders.  
}
\maketitle

\section{Introduction}
Grand Unified Theories (GUTs)\ \cite{Pati:1973rp,Pati:1974yy, Georgi:1974sy, Georgi:1974yf, Georgi:1974my, Fritzsch:1974nn} are arguably one of the most appealing extensions of the Standard Model (SM) of particle physics. In 1974, a simple and elegant GUT based on the unifying gauge group SU(5) was proposed by H.\ Georgi and S.\ Glashow\ (GG model)\ \cite{Georgi:1974sy}. However, this model is incompatible with the current experimental data for three main reasons. Firstly, the GG model does not allow for gauge coupling unification, which is a necessary condition for a GUT. Secondly, it predicts massless neutrinos, which is in conflict with neutrino oscillation experiments requiring that at least two neutrino should be massive\ \cite{SNO:2002tuh}. Thirdly, since the SM Higgs doublet is embedded into a 5-dimensional Higgs representation of SU(5), the GG model predicts the GUT scale relation between the charged lepton and down-type quark Yukawa matrices 
\begin{align}
    Y_e=Y_d^T. \label{eq:Ye=Yd}   
\end{align}
This relation in particular implies a GUT scale unification of the tau and bottom Yukawa couplings $y_\tau=y_b$, as well as a unification of the muon and strange Yukawa couplings $y_\mu=y_s$, which disagrees with the low energy data.

The first shortcoming requires extending the particle content of the minimal model by additional GUT representations and suitably splitting the masses of their component fields such that the running gauge couplings meet. 
The second shortcoming can be addressed by introducing SU(5) representations that allow neutrino mass generation at the tree level\ \cite{Dorsner:2005fq, Dorsner:2005ii, Dorsner:2006hw, Bajc:2006ia, Bajc:2007zf, FileviezPerez:2007bcw, Dorsner:2006fx} or at the loop level\ \cite{FileviezPerez:2016sal, Kumericki:2017sfc, Saad:2019vjo, Dorsner:2019vgf, Dorsner:2021qwg, Antusch:2023jok}. 
Finally, the third shortcoming can for instance be resolved by generating the Yukawa couplings from linear combinations of the renormalisable and higher dimensional non-renormalisable operators\ \cite{Ellis:1979fg}, or at the renormalizable level by either introducing a 45-dimensional Higgs field and considering linear combinations of couplings between the SM fermions and both the 5- as well as the 45-dimensional Higgs field\ \cite{Dorsner:2006dj}, or by introducing vector-like fermions which mix with the SM fermions\ \cite{Babu:2012pb, Dorsner:2014wva, FileviezPerez:2018dyf}. 

However, historically a first and very aesthetic solution for the third problem was proposed by H.\ Georgi and C.\ Jarlskog (GJ model) in 1979\ \cite{Georgi:1979df}. In their model, the particle content of the GG model is extended by a 45-dimensional Higgs field (as well as by two 5-dimensional Higgs fields). If the 45-dimensional Higgs field couples to the SM fermions this gives rise to the GUT scale relation
\begin{align}
    Y_e=-3Y_d^T. \label{eq:Ye=-3Yd} 
\end{align}
Considering a linear combination of the operators giving the relations\ \eqref{eq:Ye=Yd}\ and\ \eqref{eq:Ye=-3Yd} would, on the one hand, solve the shortcoming (as already mentioned above), but, on the other hand, predictivity in the Yukawa sector would be lost. Predictivity is however maintained if it is ensured that different generations of charged leptons and down-type quarks couple to different Higgs fields (which can, for example, be achieved when a family symmetry is introduced on top of the gauge symmetry). To achieve predictivity, without referring to any particular family symmetry, the GJ model hypothesizes the following textures of the Yukawa coupling matrices,
\begin{align}
    Y_d=
\begin{pmatrix}
0 && B && 0 \\
A && C && 0 \\
0 && 0 && D
\end{pmatrix},\quad Y_e^T=
\begin{pmatrix}
0 && B && 0 \\
A && -3C && 0 \\
0 && 0 && D
\end{pmatrix},
\end{align}
implying the GUT scale relations $y_\tau/y_b=1,\, y_\mu/y_s=-3,\, y_e/y_d=-1/3$ which were at that time compatible with the experimental data. However, the current data suggests (taking only the known SM particles into account in the renormalization group (RG) evolution) that other ratios such as $y_\tau/y_b=3/2,\, y_\mu/y_s=9/2$ are better suited (see e.g.\ \cite{Antusch:2021yqe}). 

Interestingly, these latter ratios can be obtained from higher dimensional operators\ \cite{Antusch:2009gu, Antusch:2013rxa}. With these higher dimensional operators at hand, models similar to the GJ model can be build if the following two conditions are satisfied: (i) the Yukawa matrices should be hierarchical, (ii) the 22- and 33- entry should be dominated by a single GUT operator, a concept which is referred to as \textit{single operator dominance}\ \cite{Antusch:2009gu, Antusch:2013rxa,Antusch:2019avd}.\footnote{For models in which the concept of single operator dominance has been applied, see e.g.\ \cite{Antusch:2011sq, Antusch:2011qg, Antusch:2011xz, Antusch:2012fb,Antusch:2012gv, Antusch:2013wn, Antusch:2013kna,Antusch:2013tta, Antusch:2013eca, Antusch:2014poa, Antusch:2017ano, Antusch:2017tud, Antusch:2018gnu, Antusch:2022ufb}.} 

Following this approach, non-SUSY GUT scenarios in which neutrino masses are generated by a type I or a type II seesaw have been investigated in\ \cite{Antusch:2021yqe}, respectively \cite{Antusch:2022afk}. For GUT scenarios with a type I seesaw it was shown that the GUT scale ratios $y_\tau/y_b=3/2$ and $y_\mu/y_s=9/2$ are compatible with the experimental data. Moreover, for GUT scenarios in which neutrino masses are generated by a type II seesaw it was found, that two combinations of GUT scale relations are viable, namely (i) $y_\tau/y_b=3/2$ and $y_\mu/y_s=9/2$ and (ii) $y_\tau/y_b=2$ and $y_\mu/y_s=6$. 

In this paper we will investigate the viability of such GUT scale ratios for the case that neutrino masses stem from a combination of a type I\ \cite{Minkowski:1977sc, Yanagida:1979as, Gell-Mann:1979vob, Glashow:1979nm, Mohapatra:1980yp} and a type III\ \cite{Foot:1988aq} seesaw mechanism. In this regard, we will consider a GUT scenario in which the particle content of the GG model is extended by a fermionic adjoint representation as well as by a 45-dimensional Higgs field.\footnote{A non-supersymmetric SU(5) GUT with this particle content was first considered in\ \cite{FileviezPerez:2007bcw}. However, so far it has not been studied under the assumption of \textit{single operator dominance}.} The former representation is needed to generate neutrino masses, while the latter gives rise to operators yielding potentially viable GUT scale Yukawa ratios. Moreover, both of these representations help to allow for gauge coupling unification. Using the Mathematica package \texttt{ProtonDecay}\ \cite{Antusch:2020ztu} and extending the above scenario to ``toy models'' we also compute the nucleon decay widths for various decay channels. Finally, we compute the masses of the added fermion and scalar fields. 

The paper is organized as follows: While the GUT scenario as well as the toy models are introduced in Section\ \ref{sec:GUT scenario}, the procedure for the numerical analysis is explained in Section\ \ref{sec:numerical procedure}. In Section\ \ref{sec:Results} the results are presented and discussed, before concluding in Section\ \ref{sec:conclusion}. In Appendix\ \ref{sec:AppendixA}, definitions of the newly introduced Yukawa couplings are given, while all relevant RGEs that we have derived are listed in Appendix\ \ref{sec:AppendixB}.

\section{GUT scenario}\label{sec:GUT scenario}

\subsection{Particle content}
The SM fermions are embedded as usual into three generations of ${\bm{\overline{5}_F}}_i$ and ${\bm{10_F}}_i$
\begin{align}
{\bm{\overline{5}_F}}_i=&\;d_i^c(\overline{3},1,\frac{1}{3})\oplus \ell_i(1,2,-\frac{1}{2}),\\
{\bm{10_F}}_i=&\; q_i (3,2,\frac{1}{6}) \oplus u_i^c (\overline{3},1,-\frac{2}{3}) \oplus e_i^c (1,1,1).
\end{align}
In the considered scenario, neutrino masses are generated via a combination of a type I and a type III seesaw mechanism. The corresponding fermionic singlet $\Sigma_c$ and triplet $\Sigma_b$ (under $SU(2)_L$) are contained in an adjoint fermionic representation
\begin{align}
\bm{24_F}=&\;\Sigma_a(8,1,0)\oplus \Sigma_b(1,3,0)\oplus\Sigma_c(1,1,0)\oplus \Sigma_d(3,2,-\frac{5}{6})\oplus\Sigma_e(\overline{3},2,\frac{5}{6}).    
\end{align}
Moreover, the GUT Higgs fields decompose under the SM gauge group as
\begin{align}
\bm{24_H}=&\;\Phi_a(8,1,0)\oplus \Phi_b(1,3,0)\oplus\Phi_c(1,1,0)\oplus \Phi_d(3,2,-\frac{5}{6})\oplus\Phi_e(\overline{3},2,\frac{5}{6}),\\
\bm{5_H}=&\;T_a(3,1,-\frac{1}{3})\oplus H_a(1,2,\frac{1}{2}),\\
\bm{45_H}=&\;\phi_a(8,2,\frac{1}{2})\oplus\phi_b(6,1,-\frac{1}{3})\oplus\phi_c(3,3,-\frac{1}{3})\oplus\phi_d(\overline{3},2,-\frac{7}{6})\oplus\phi_e(\overline{3},1,-\frac{4}{3})\nonumber\\
&\oplus T_b(3,1,-\frac{1}{3})\oplus H_b (1,2,\frac{1}{2}).
\end{align}
After the SU(5) breaking, the color triplets $T_a$ and $T_b$ mix to yield the mass eigenstates $t_1=\cos(\alpha)T_a+\sin(\alpha)T_b$ and $t_2 =-\sin(\alpha)T_a+\cos(\alpha)T_b$. Similarly, $H_a$ and $H_b$ mix to form the mass eigenstates $h_1=\cos(\beta)H_a+\sin(\beta)H_b$ and $h_2^\perp=-\sin(\beta)H_a+\cos(\beta)H_b$, where $h_1$ is the SM Higgs doublet.

\subsection{Neutrino masses}
At tree-level the relevant GUT operators for neutrino mass generation read\footnote{After the GUT symmetry breaking these two GUT operators decompose into 19 SM Yukawa interactions. For details see Appendix\ \ref{sec:AppendixA}.}
\begin{align}
\mathcal{L}\supset Y_A\,\bm{\overline{5}_F} \bm{24_F}\bm{5_H} +Y_B\,\bm{\overline{5}_F} \bm{24_F}\bm{45_H}.
\end{align}
After the GUT symmetry breaking the following relevant terms emerge
\begin{align}
    \mathcal{L}\supset -Y_2 \ell \Sigma_b H_a- Y_8 \ell \Sigma_b H_b -Y_4 \ell \Sigma_c H_a -Y_{13} \ell \Sigma_c H_b - m_{\Sigma_b}\Sigma_b\Sigma_b - m_{\Sigma_c}\Sigma_c\Sigma_c,
\end{align}
where $m_{\Sigma_b}$ and $m_{\Sigma_b}$ are the respective masses of $\Sigma_b$ and $\Sigma_c$, and where the GUT scale relations 
\begin{align}
    Y_2=-\sqrt{\frac{3}{10}}\,Y_A,\quad Y_4=Y_A,\quad Y_8=\frac{\sqrt{5}}{4}\,Y_B,\quad \text{and} \quad Y_{13}=\frac{\sqrt{3}}{4}\,Y_B
\end{align} 
hold.
After the SU(2) triplet $\Sigma_b$ and SU(2) singlet $\Sigma_c$ have been integrated out and the two Higgs fields $H_a$ and $H_b$ have taken their vacuum expectation values (vevs) $v_a$ and $v_b$, where $v_a^2+v_b^2=v^2=(246\ \text{GeV})^2$, and where $v_a=v\cos(\beta)$ and $v_b=v\sin(\beta)$, the neutrino mass matrix $m_\nu$ reads
\begin{align}
    m_{\nu}^{ij}=-\frac{(Y_2^i\,v_a+Y_8^i\,v_b)(Y_2^j\,v_a+Y_8^j\,v_b)}{4m_{\Sigma_b}}-\frac{(Y_4^i\,v_a+Y_{13}^i\,v_b)(Y_4^j\,v_a+Y_{13}^j\,v_b)}{4m_{\Sigma_c}}\,.
\end{align}
Since the neutrino mass matrix $m_\nu$ is of rank two, two massive and one massless neutrino are predicted.

\subsection{Quark-lepton Yukawa ratios}
With $\bm{X}$ and $\bm{Y}$ representing one or multiple Higgs fields, the charged fermion masses stem from GUT operators of the form 
\begin{align}
    \bm{Y_{\overline{5}}}^{ij}\,:\quad {\bm{10_{F}}}_i{\bm{\overline{5}_F}}_j\bm{X}\quad&\supset\quad Y_d^{ij},\ Y_e^{ij}\\
    \bm{Y_{10}}^{ij}\,:\quad {\bm{10_{F}}}_i{\bm{10_{F}}}_j\bm{Y}\quad&\supset\quad Y_u^{ij},
\end{align}
where $Y_u$, $Y_d$ and $Y_e$ denote the usual SM charged fermion Yukawa matrices.
Assuming in the charged fermion Yukawa sector the concept of \textit{single operator dominance}, i.e. that each Yukawa entry is dominated by a singlet GUT operator, allows to connect the down-type with the charged lepton Yukawa matrix via group theoretical Clebsch-Gordan (CG) factors $c_{ij}$.
In SU(5) GUTs, and considering up to dimension five operators, the potentially viable CG factors are $|c_{ij}|\in\lbrace1/6,\, 1/2,\, 2/3,\, 1,\, 3/2,\, 2,\, 3,\, 9/2,\, 6,\, 9,\, 18\rbrace$. The possible GUT operators yielding these ratios are given in \cite{Antusch:2009gu, Antusch:2013rxa}. 
Moreover, if the matrix $\bm{Y_{\overline{5}}}$ is assumed to be of hierarchical nature and dominated by its diagonal entries, then the second and third family down-type quark and charged lepton masses stem dominantly from the GUT operators $\mathcal{O}_2$ and $\mathcal{O}_3$ dominating the 22 and 33 positions in $\bm{Y_{\overline{5}}}$. Depending on which operators are chosen for $\mathcal{O}_2$ and $\mathcal{O}_3$, different GUT scale Yukawa ratios $y_\tau/y_b$ and $y_\mu/y_s$ are predicted. Our numerical analysis (cf.\ Section\ \ref{sec:Results}) shows that there are only two possible choices for the GUT scale ratio $y_\tau/y_b$, namely 3/2 or 2. The former CG factor can be complemented by a factor 9/2 for the second family, while for the latter CG factor two different completions, $y_\mu/y_s=6$ or $y_\mu/y_s=9/2$, are possible. 

\subsection{Toy models}\label{sec:Toy models}
We now extend the above motivated scenarios to three toy models which also include the first family. For simplicity, we chose the matrix $\bm{Y_{\overline{5}}}$ to be of diagonal nature. The double ratio $(y_\mu y_d)/(y_ey_s)=10.7^{+1.6}_{-0.9}$, which is nearly constant under renormalization group running (see e.g.\ \cite{Antusch:2013jca}), suggests, that the the ratio $y_\mu/y_s=9/2$ is best complemented by a ratio $y_e/y_d=4/9$, while the best completion of the ratio $y_\mu/y_s=6$ is given by $y_e/y_d=1/2$. Utilizing these ratios, our three toy models relate the down-type with the charged lepton Yukawa matrix via 
\begin{align}
&\text{Model 1:}\quad Y_e=\text{diag}\left(\frac{4}{9},\frac{9}{2},\frac{3}{2}\right)\cdot Y_d^T\,,\label{eq:model1}\\
&\text{Model 2:}\quad Y_e=\text{diag}\left(\frac{4}{9},\frac{9}{2},2\right)\cdot Y_d^T\,,\label{eq:model2}\\
&\text{Model 3:}\quad Y_e=\text{diag}\left(\frac{1}{2},6,2\right)\cdot Y_d^T\,.  \label{eq:model3}  
\end{align}

Moreover, for simplicity\footnote{We might consider higher-dimensional operators also for $\bm{Y_{10}}$, for example to explain the mass hierarchy, however since no Yukawa ratio predictions arise from this sector, we stick to the simplest case in our toy models.} we assume in each toy model that $\bm{Y_{10}}$ is dominated by the operator $\bm{10_F10_F5_H}$ in all entries, yielding a symmetric up-type Yukawa matrix, i.e. $Y_u=Y_u^T$. Finally, in all toy models neutrino masses stem from a linear combination of the operators $\bm{\overline{5}_F24_F5_H}$ and $\bm{\overline{5}_F24_F45_H}$.

\section{Numerical procedure}\label{sec:numerical procedure}

\subsection{Implementation of the charged fermion Yukawa sector}
We implement all three toy models at the GUT scale as described in Section\ \ref{sec:Toy models}. In all three models the down-type Yukawa matrix $Y_d$ is simply implemented as 
\begin{align}
Y_d=\text{diag}(y_1^d,y_2^d,y_3^d), 
\end{align}
while the charged lepton Yukawa matrix $Y_e$ is implemented according to Eq.\ \eqref{eq:model1},\ \eqref{eq:model2},\ and\ \eqref{eq:model3}, respectively. Since $Y_u$ is symmetric we use a Takagi decomposition and implement it as 
\begin{align}
Y_u = U_{u}^\dagger Y_u^\text{diag} U_{u}^*,
\end{align} 
where\footnote{Here we have dropped three unphysical parameters but kept the GUT phases $\beta_1^u$ and $\beta_2^u$ which effect the nucleon decay widths\ \cite{Ellis:1979hy,Ellis:2019fwf}.}
\begin{align}
    U_u =
    \begin{pmatrix}
    1 & 0 & 0 \\
    0 & c_{23}^u & s_{23}^u \\
    0 & -s_{23}^u & c_{23}^u
    \end{pmatrix}
    \begin{pmatrix}
    c_{13}^u & 0 & s_{13}^ue^{-i \delta^u}  \\
    0 & 1 & 0 \\
    -s_{13}^ue^{i \delta^u} & 0 & c_{13}^u  
    \end{pmatrix}
    \begin{pmatrix}
    c_{12}^u & s_{12}^u & 0 \\
    -s_{12}^u & c_{12}^u & 0 \\
    0 & 0 & 1
    \end{pmatrix}
    \begin{pmatrix}
    e^{i\beta_1^u}&0&0\\
    0&e^{i\beta_2^u}&0\\
    0&0&1
    \end{pmatrix},
\end{align}
and where $Y_u^\text{diag}=\text{diag}(y_1^u,y_2^u,y_3^u)$.

\subsection{Implementation of the neutrino sector}
In order to simplify the analysis we assume in the neutrino sector that the Yukawa matrices $Y_{5}$ and $Y_{6}$ (for the definitions of these couplings, see Appendix\ \ref{sec:AppendixA}) are of the form
\begin{align}\label{eq:CSD3}
    Y_{5}=z_1 \begin{pmatrix}
    0\\1\\1
    \end{pmatrix},\quad\quad
    Y_{6}=z_2 \begin{pmatrix}
    1\\1\\3
    \end{pmatrix},
\end{align}
where $z_1$ and $z_2$ are real parameters. Furthermore, we denote the relative phase difference between $m_{\Sigma_b}$ and $m_{\Sigma_c}$ by $\gamma$ (i.e. $\gamma=\arg(m_{\Sigma_b}/m_{\Sigma_c})$). This structure is motivated by CSD3\ \cite{King:2013iva} which in the case of type I seesaw has been shown to correctly describe the low-scale neutrino observables together with a normal neutrino mass hierarchy (see e.g.\ \cite{Antusch:2021yqe} for a recent work). 

\subsection{GUT scale parameters and low energy observables}
Each toy model contains 33 input parameters which decompose into the GUT scale $M_{\text{GUT}}$, the SU(5) gauge coupling $g_{\text{GUT}}$, the masses of the added particles,\footnote{Note that $m_{\Sigma_d}=m_{\Sigma_e}$.} $m_{\Phi_a}$, $m_{\Phi_b}$, $m_{\phi_a}$, $m_{\phi_b}$, $m_{\phi_c}$, $m_{\phi_d}$, $m_{\phi_e}$, $m_{\Sigma_a}$, $m_{\Sigma_b}$, $m_{\Sigma_c}$, $m_{\Sigma_d}$, $m_{t_1}$, $m_{t_2}$, $m_{h_2}$,  the singular values $y_1^u$, $y_2^u$, $y_3^u$, $y_1^d$, $y_2^d$, $y_3^d$ and angles $\theta_{12}^u$, $\theta_{13}^u$, $\theta_{23}^u$ as well as phases $\delta^u$, $\beta_1^u$, $\beta_2^u$ of the charged fermion Yukawa matrices, the parameters of the neutrino Yukawa couplings $z_1$, $z_2$, and $\gamma$, and the eigenstate mixing angles $\alpha$ and $\beta$. The respective ranges of these input parameters are given by\footnote{Note that although we do not put any perturbativity constraints on the neutrino Yukawa couplings $z_1$ and $z_2$ the fit automatically choses them to be below 1 (cf.\ Section\ \ref{sec:Results}).}
\begin{align}\label{eq:input parameters}
    M_\text{GUT}&< M_{Pl},  \nonumber\\ 
    m_{\Phi_a}, m_{\Phi_b}, m_{\phi_a}, m_{\phi_b}, m_{\phi_c}, m_{\phi_d}, m_{\phi_e}, m_{\Sigma_a}, m_{\Sigma_b}, m_{\Sigma_c}, m_{\Sigma_d}, m_{h_2} &\in [1\,\text{TeV},M_\text{GUT}],  \nonumber\\
m_{t_1},m_{t_2}&\in [10^{11}\,\text{GeV},M_{\text{GUT}}], \nonumber\\
g_{\text{GUT}},\,y_1^u,\,y_2^u,\,y_3^u,\,y_1^d,\,y_2^d,\,y_3^d&\in[0,1],   \\
\theta_{12}^u,\,\theta_{13}^u,\,\theta_{23}^u,\,\alpha,\,\beta &\in[0,\pi/2],   \nonumber\\
\delta^u,\,\beta_1^u,\,\beta_2^u,\,\gamma &\in[-\pi,\pi), \nonumber\\
z_1,\,z_2 &>0. \nonumber
\end{align}
These input parameters are fitted to the 22 low-scale observables (listed in Eq.\ \eqref{eq:observables}) and the nucleon decay widths of thirteen decay channels (listed in Table\ \ref{tab:nucleon decay}). 
\begin{align}\label{eq:observables}
&g_1,\,g_2,\,g_3,\nonumber\\
&y_u,\,y_c,\,y_t,\,y_d,\,y_s,\,y_b,\,\theta_{12}^{\text{CKM}},\,\theta_{13}^{\text{CKM}},\,\theta_{23}^{\text{CKM}},\,\delta^{\text{CKM}},\,y_e,\,y_\mu,\,y_\tau,\,\hspace{0.6cm}\\
&\Delta m_{21}^2,\,\Delta m_{31}^2,\,\theta_{12}^{\text{PMNS}},\,\theta_{13}^{\text{PMNS}},\,\theta_{23}^{\text{PMNS}},\,\delta^{\text{PMNS}}.\nonumber
\end{align}
For the SM gauge couplings and Yukawa observables we take the experimental values from\ \cite{Antusch:2013jca}, while the values for the neutrino sector are taken from\ NuFIT\ 5.1\ \cite{Esteban:2020cvm}.

\begin{table}[t!]
\centering
\begin{tabu}{llllc}\tabucline[1.1pt]{-} \noalign{\vskip 2mm}
& decay channel  &  $\tau/\mathcal{B}$ [year] & $\Gamma_{\text{partial}}$ [GeV] & Reference  
\\\noalign{\vskip 1mm}\hline\noalign{\vskip 2mm}
Proton: & $p\rightarrow \pi^0\,e^+$  &  $>\,2.4\cdot 10^{34}$  &  $<\,8.7\cdot 10^{-67}$ & \cite{Super-Kamiokande:2020wjk}  \\\noalign{\vskip 2mm}
&  $p\rightarrow \pi^0\,\mu^+$  &  $>\,1.6\cdot 10^{34}$  &  $<\,1.3\cdot 10^{-66}$ & \cite{Super-Kamiokande:2020wjk}  \\\noalign{\vskip 2mm}
&  $p\rightarrow \eta^0\,e^+$  &  $>\,1.0\cdot 10^{34}$  &  $<\,2.0\cdot 10^{-66}$ & \cite{Super-Kamiokande:2017gev}  \\\noalign{\vskip 2mm}
&  $p\rightarrow \eta^0\,\mu^+$  &  $>\,4.7\cdot 10^{33}$  &  $<\,4.4\cdot 10^{-66}$ & \cite{Super-Kamiokande:2017gev}   \\\noalign{\vskip 2mm}
&  $p\rightarrow K^0\,e^+$  &  $>\,1.1\cdot 10^{33}$  &  $<\,1.9\cdot 10^{-65}$ & \cite{Brock:2012ogj}  \\ \noalign{\vskip 2mm}
&  $p\rightarrow K^0\,\mu^+$  &  $>\,3.6\cdot 10^{33}$  &  $<\, 5.8\cdot 10^{-66}$ & \cite{Super-Kamiokande:2022egr}  \\\noalign{\vskip 2mm}
&  $p\rightarrow \pi^+\,\overline{\nu}$  &  $>\,3.9\cdot 10^{32}$  &  $<\,5.3\cdot 10^{-65}$ & \cite{Super-Kamiokande:2013rwg}  \\\noalign{\vskip 2mm}
&  $p\rightarrow K^+\,\overline{\nu}$  &  $>\,6.6\cdot 10^{33}$  &  $<\,3.2\cdot 10^{-66}$ & \cite{Takhistov:2016eqm} 
\\\noalign{\vskip 1mm}\hline\noalign{\vskip 2mm}
Neutron:  &  $n\rightarrow \pi^-\,e^+$  &  $>\,5.3\cdot 10^{33}$  &  $<\,3.9\cdot 10^{-66}$ & \cite{Super-Kamiokande:2017gev}   \\\noalign{\vskip 2mm}  
&  $n\rightarrow \pi^-\,\mu^+$  &  $>\,3.5\cdot 10^{33}$  &  $<\,5.9\cdot 10^{-66}$ & \cite{Super-Kamiokande:2017gev}   \\\noalign{\vskip 2mm}
&  $n\rightarrow \pi^0\,\overline{\nu}$  &  $>\,1.1\cdot 10^{33}$  &  $<\,1.9\cdot 10^{-65}$ & \cite{Super-Kamiokande:2013rwg}  \\\noalign{\vskip 2mm}
&  $n\rightarrow \eta^0\,\overline{\nu}$  &  $>\,5.6\cdot 10^{32}$  &  $<\,3.7\cdot 10^{-65}$ & \cite{Brock:2012ogj}  \\\noalign{\vskip 2mm}
&  $n\rightarrow K^0\,\overline{\nu}$  &  $>\,1.2\cdot 10^{32}$  &  $<\,1.7\cdot 10^{-64}$ & \cite{Brock:2012ogj}  \\\noalign{\vskip 1mm}
\tabucline[1.1pt]{-}
\end{tabu}
\caption{Current experimental bounds on the decay widths $\Gamma_{\text{partial}}$, respectively lifetime $\tau/\mathcal{B} $ at 90\ \% confidence level, where $\mathcal{B}$ is the branching ratio for the decay channel. See also~\cite{Dev:2022jbf} for future projections and sensitivities of various upcoming detectors. }\label{tab:nucleon decay}
\end{table}

\subsection{Fitting procedure}\label{sec:fitting procedure}
After implementing the input parameters given in Eq.\ \eqref{eq:input parameters} at the GUT scale we compute the RG evolution to the $Z$ scale. For the gauge couplings we use a 2-loop running, while we compute the running of the Yukawa matrices and the effective neutrino mass operator at 1-loop. The nucleon decay widths are computed using the Mathematica package \texttt{Proton Decay}\ \cite{Antusch:2020ztu} (for a description of the calculation see e.g.\ \cite{Antusch:2021yqe}). Taking into account all observables we compute at the low scale the $\chi^2$-function which we minimize using a differential evolution algorithm giving us a benchmark point. With a flat prior distribution we calculate $4\times 10^6$ data points performing a Markov-chain-Monte-Carlo (MCMC) analysis using an adaptive Metropolis-Hastings algorithm\ \cite{Metropolis-Hastings-algorithm} which we start from this benchmark point. These data points are finally used to compute the highest posterior density (HPD) ranges of various quantities.

\section{Results}\label{sec:Results}
The results of our numerical analysis are presented in this section. We are in particular interested in the nucleon decay predictions, the intermediate-scale particle masses as well as the low scale predictions for the charged lepton and down-type quark mass ratios. In Section\ \ref{sec:benchmark points} we show the results of our minimization procedure. Starting an MCMC analysis from these benchmark points allows us to obtain the HPD ranges of various quantities. The results of this analysis is presented in Section\ \ref{sec:highest posterior densities}.

\subsection{Benchmark points}\label{sec:benchmark points}
We obtain for all three models benchmark points through a minimization of the $\chi^2$-function as described in Section\ \ref{sec:numerical procedure}. In Table\ \ref{tab-input parameters benchmark points} the input parameters for the respective benchmark points are listed. Moreover, the dominant pulls $\chi_i^2$ are presented in Table\ \ref{tab-chi squared benchmark points}. All three models can be very well fitted to the data. The strongest (though quite small) pull is given by the first and second family down-type quark masses. The biggest difference between the three models is the respectively favored GUT scale. For Models 2 and 3 a GUT scale above $10^{17}$\ GeV is favored, while for the benchmark point of Model 1 a GUT scale below $10^{16}$\ GeV is obtained. This also results in different results for the predicted nucleon decay rates (cf.\ Section\ \ref{sec:highest posterior densities}). Another difference is the preferred choice of some of the intermediate-scale particle masses. In the presented benchmark point the mass of the fermionic field $\Sigma_a$ is obtained to be at the GUT scale for Model 1, at the intermediate scale for Model 2 and at the relatively low scale (23\ TeV) for Model 3. Moreover, a mass of the leptoquark $\phi_c$ of 1\ TeV, respectively 4\ TeV is obtained for Model 3, respectively Model 2, whereas for Model 1 the mass of this field is above $10^6$\ TeV. For the HPD results of these particle masses confer the subsequent section.

\begin{table}[ht]
\centering
\renewcommand{\arraystretch}{1.1}
\begin{tabu}{|c|ccc|ccc|ccc|}
\tabucline[1.1pt]{-}
&&Model 1&&&Model 2&&&Model 3&\\
\hline
$g_\text{GUT}\;/\;10^{-1}$ & & 5.94 & & & 6.17 & & & 6.33 & \\ 
$\log_{10}(M_{\text{GUT}}\;/\;\text{GeV})$ & & 15.6 & & & 17.2 & & & 17.3 &\\
$\log_{10}(m_{\phi_a}\;/\;\text{GeV})$ & & 9.43 & & & 14.0 & & & 16.7 &\\ 
$\log_{10}(m_{\phi_c}\;/\;\text{GeV})$ & & 9.02 & & & 3.63 & & & 3.00 &\\ 
$\log_{10}(m_{\Sigma_a}\;/\;\text{GeV})$ & & 15.6 & & & 7.53 & & & 4.36 &\\
$\log_{10}(m_{\Sigma_b}\;/\;\text{GeV})$ & & 14.2 & & & 14.9 & & & 14.7 &\\
$\log_{10}(m_{\Sigma_c}\;/\;\text{GeV})$ & & 13.8 & & & 12.8 & & & 13.2 &\\
$\log_{10}(m_{\Sigma_d}\;/\;\text{GeV})$ & & 14.2 & & & 15.9 & & & 14.1 &\\
$y_1^u\;/\;10^{-6}$ & & 2.63 & & & 2.11 & & & 1.99 &\\
$y_2^u\;/\;10^{-3}$ & & 1.46 & & & 1.37 & & & 1.18 &\\
$y_3^u\;/\;10^{-1}$ & & 4.54 & & & 4.26 & & & 3.65 &\\
$y_1^d\;/\;10^{-6}$ & & 6.21 & & & 6.30 & & & 5.46 & \\ 
$y_2^d\;/\;10^{-4}$ & & 1.31 & & & 1.21 & & & 0.99 & \\ 
$y_3^d\;/\;10^{-3}$ & & 6.64 & & & 6.01 & & & 5.36 & \\ 
$z_1\;/\;10^{-1}$ & & 3.50 & & & 9.42 & & & 6.86 & \\ 
$z_2\;/\;10^{-1}$ & & 1.12 & & & 0.32 & & & 0.50 & \\
$\gamma$ & & 1.85 & & & 1.48 & & & 1.68 & \\
$\alpha$ & & 0.50 & & & 1.00 & & & 0.50 & \\
\tabucline[1.1pt]{-}
\end{tabu} 
\caption{The GUT scale input parameters of the benchmark points for all three models.}\label{tab-input parameters benchmark points}
\end{table}

\begin{table}[ht]
\centering
\renewcommand{\arraystretch}{1.1}
\begin{tabu}{|c|ccccccc|}
\tabucline[1.1pt]{-}
& $\chi^2$ & $\chi^2_{y_d}$ & $\chi^2_{y_s}$ & $\chi^2_{y_b}$ & $\chi^2_{y_\mu} $& $\chi^2_{y_\tau}$ & $\chi^2_{\Gamma({p\rightarrow \pi^0e^+})}$ \\
\hline
Model 1 & 1.36 & 0.27 & 0.41 & 0.06 & 0.04 & 0.14 & 0.44 \\
Model 2 & 0.31 & 0.23 & 0.02 & 0.01 & 0.00 & 0.05 & 0.00 \\
Model 3 & 0.33 & 0.17 & 0.03 & 0.00 & 0.06 & 0.07 & 0.00 \\
\tabucline[1.1pt]{-}
\end{tabu} 
\caption{The total $\chi^2$ as well as the dominant pulls $\chi_i^2$ for the benchmark points of all three models.}\label{tab-chi squared benchmark points}
\end{table}

\subsection{Highest posterior densities}\label{sec:highest posterior densities}
As described in Section\ \ref{sec:fitting procedure} we vary the input parameters listed in\ Eq.\ \ref{eq:input parameters} around their benchmark points (cf.\ Table\ \ref{tab-input parameters benchmark points}) using an MCMC analysis. From these generated points we then compute the HPD intervals of various parameters and observables. 

In Figures\ \ref{fig:HPD Yukawa}\ --\ \ref{fig:decaywidhts} we use the following color coding: For Model 1, 2, and 3 the HPD intervals of various quantities are colored red, green, and blue, respectively, while the 1$\sigma$ (2$\sigma$) HPD intervals are colored dark (light).

\subsubsection{Quark-lepton mass ratios}
\begin{figure}
    \centering
    \includegraphics[width=4.3cm]{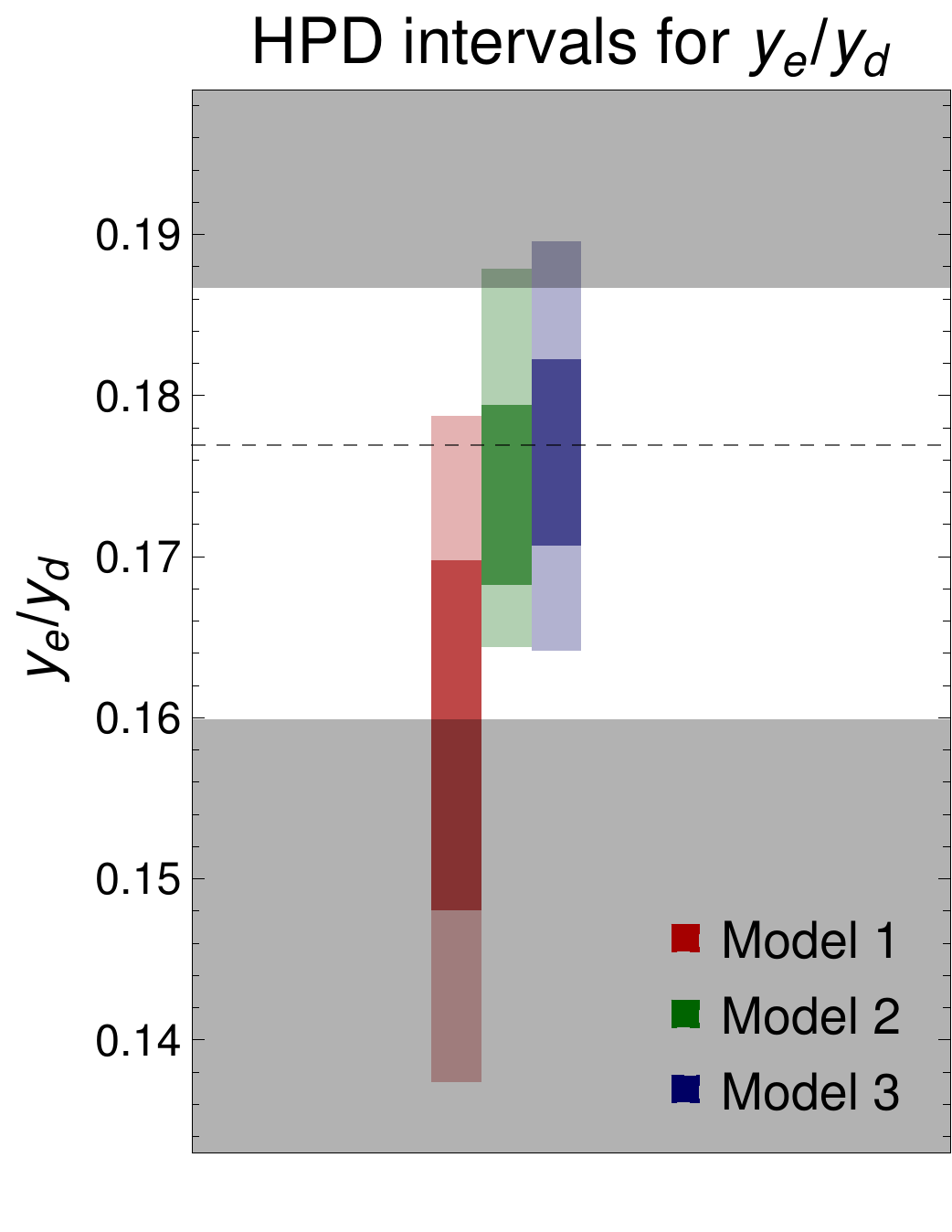}\hspace{2mm}
    \includegraphics[width=4.3cm]{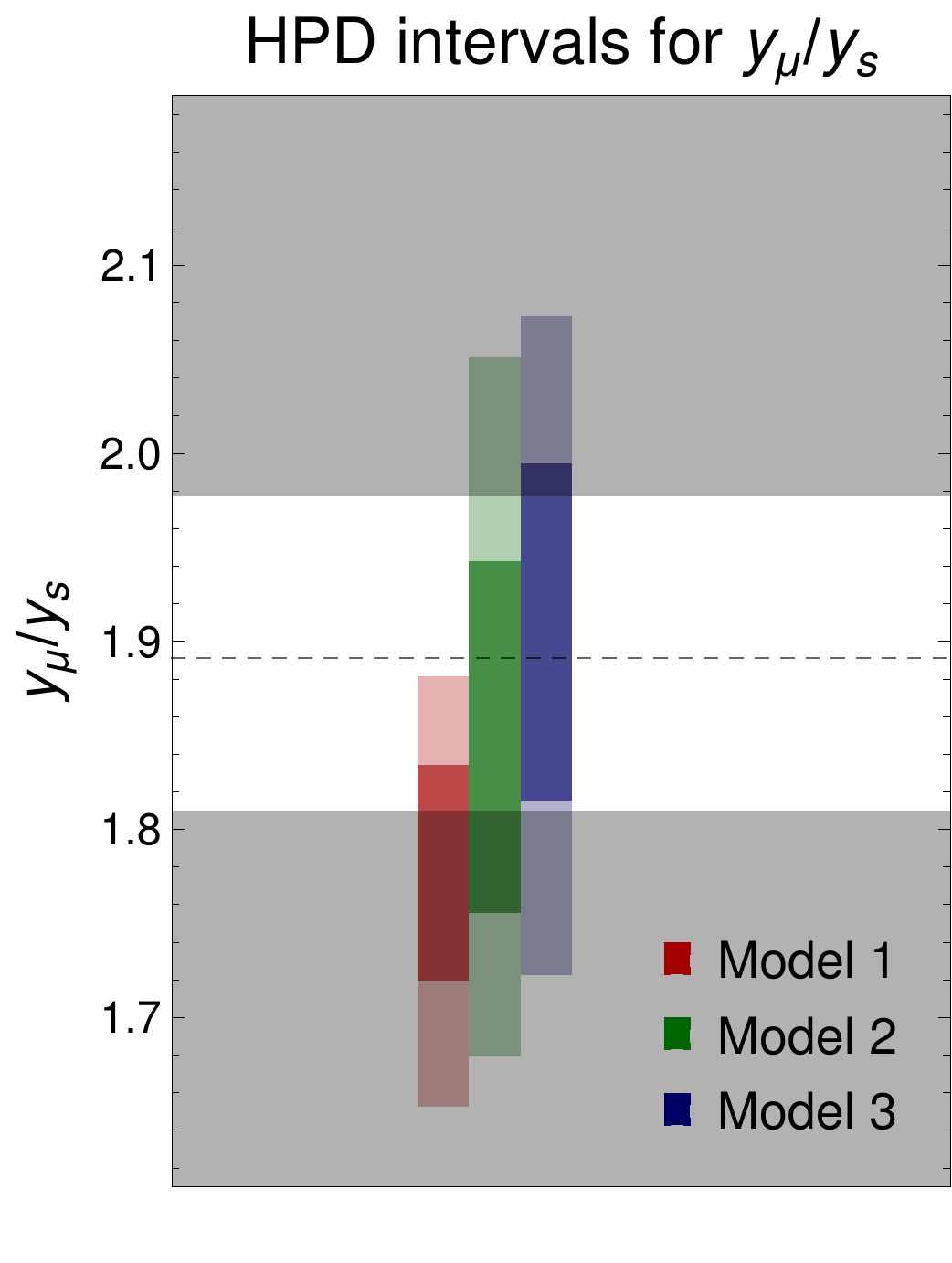}\hspace{2mm}
    \includegraphics[width=4.3cm]{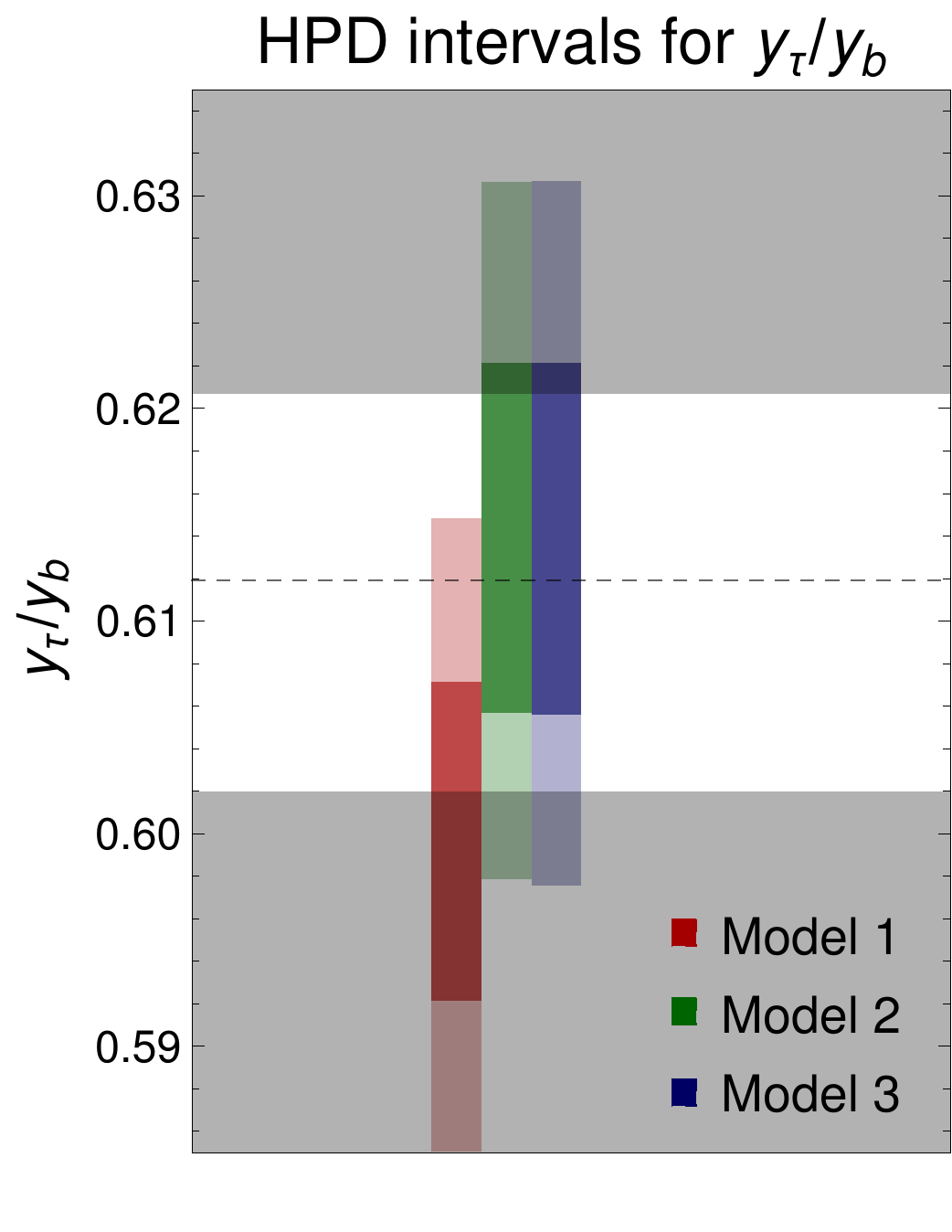}
    \caption{Low scale ($M_Z$) HPD intervals for charged lepton and down-type quark Yukawa ratios of all three families. The 1$\sigma$ (2$\sigma$) HDP intervals are colored dark (light).}
    \label{fig:HPD Yukawa}
\end{figure}

The HPD results for the low scale charged lepton and down-type quark mass ratios are presented in Figure\ \ref{fig:HPD Yukawa}. The horizontal dashed line represents the current experimental central value, whereas the white region shows the current experimental 1$\sigma$ range. Clearly, all three models are capable of reproducing viable mass ratios. This strengthens the results of the benchmark points in the previous subsection (cf.\ Tables\ \ref{tab-input parameters benchmark points}\ and\ \ref{tab-chi squared benchmark points}). Compared to Model 2 and 3, Model 1 gives a bit smaller predictions for the mass ratios for all three generations. 

\begin{figure}[t]
    \centering
    \includegraphics[width=\textwidth]{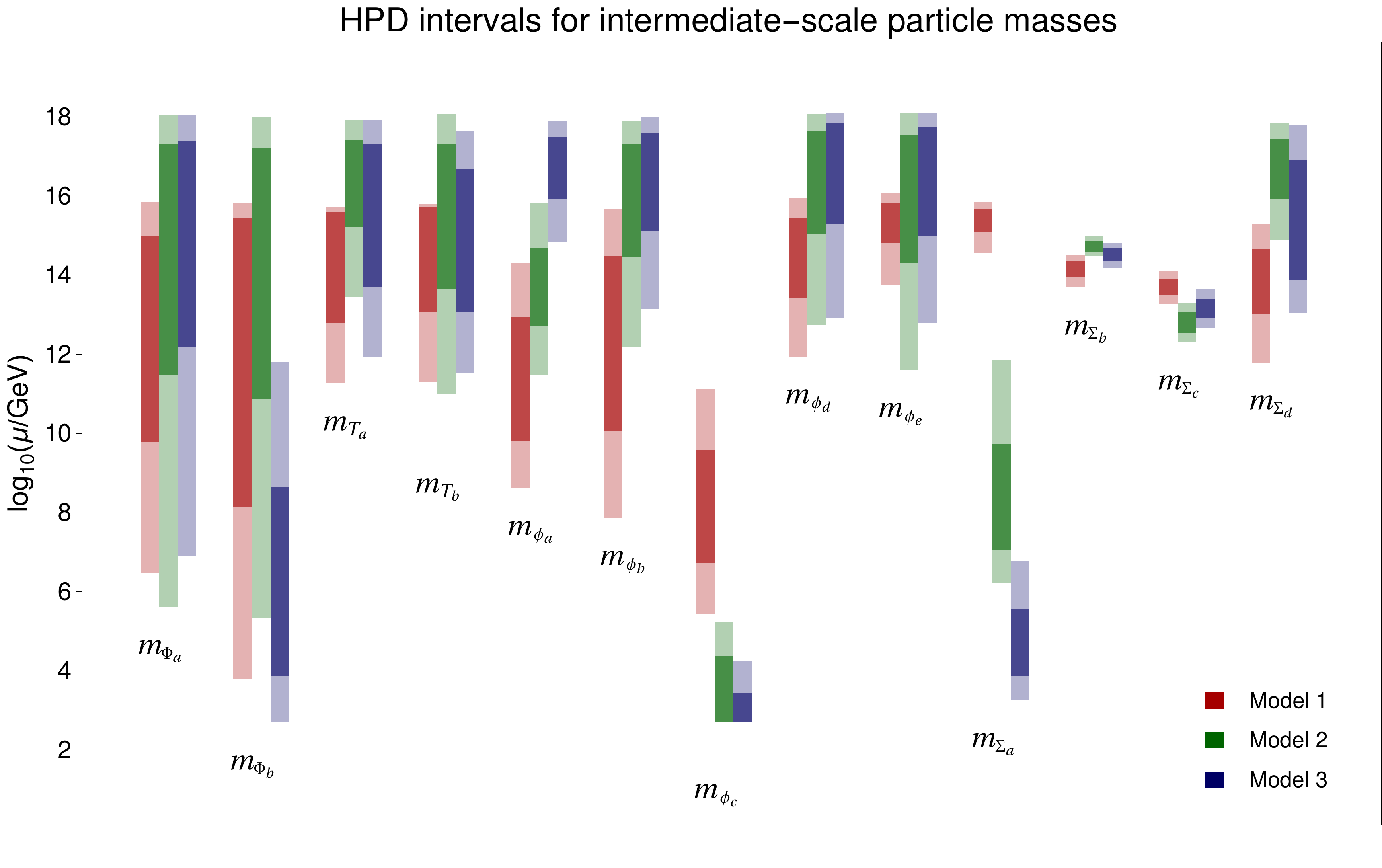}
    \caption{HPD intervals of the intermediate-scale particle masses. The 1$\sigma$ (2$\sigma$) HDP intervals are colored dark (light).}
    \label{fig:masses}
\end{figure}
\subsubsection{Intermediate-scale particle masses}
Figure\ \ref{fig:masses} shows the predicted HPD intervals of the intermediate-scale particle masses. Most of the masses are predicted to be out of the reach of current and future colliders, because they would either produce too much proton decay, spoil gauge coupling unification or because of the fit of the fermion masses. But interestingly, the fields $\Phi_b$, $\phi_c$ and $\Sigma_a$ are not only potentially within the reach of future searches, but can also be used to distinguish between the different models: An observation of the one of the fields $\Phi_b$ or $\Sigma_a$ would strongly hint towards Model 3, while an observation of the field $\phi_c$ would disfavor Model 1. In fact, the most promising lookout could be for the leptoquark $\phi_c$. The upper bound of the HPD 1$\sigma$ range is predicted to be 23\ TeV (2.8\ TeV) in Model 2 (3), whereas the upper bound of the 2$\sigma$ intervals is 175\ TeV (17\ TeV). In the following, we briefly state the current collider bounds on these particles.

The scalar triplet, $\Phi_b$, with zero hypercharge,  residing in the $\bm{24_H}$ multiplet is expected to be light in Model 3. Note that  $\Phi_b$ contains a neutral  $\Phi_b^0$ and a pair of singly charged  $\Phi_b^\pm$ states. In the low-energy effective theory, a term of the form $h_1^\dagger\Phi_b^2 h_1$ is allowed, where $h_1$ is the SM Higgs doublet. As a result of this coupling, the SM Higgs can decay into two photons $h^0\to \gamma\gamma$ via a one-loop diagram mediated by the $\Phi_b^\pm$ states. Consistency with the LHC data requires these charged states to have masses above 250 GeV \cite{Chabab:2018ert}.

The scalar leptoquark $\phi_c$, which is a triplet of SU(2)$_L$, resides around the TeV scale in  Models 2 and 3. In both models, its coupling to the SM fermions is dominated by the third-generation quark and lepton. Hence, within our scenarios, its decay branching fraction is dominated by a $b\tau$ final state. Since leptoquarks carry color, they are efficiently produced at the LHC through gluon-initiated as well as quark-initiated processes~\cite{Diaz:2017lit}. LHC searches of $pp\to b\overline b\tau\overline\tau$ from pair-produced leptoquarks rule out leptoquark masses below 1400 GeV~\cite{ATLAS:2019qpq,ATLAS:2021oiz}.

As can be seen from Eq.~\eqref{decomposed}, the color octet fermion $\Sigma_a$, which is expected to be light in Model 3, couples, for example, to a singlet down-quark  (lepton doublet) and a super-heavy colored triplet (octet) scalar. Consequently, the lifetime of a TeV scale $\Sigma_a$ is expected to be large, and it behaves like a long-lived gluino that typically arises in  split-supersymmetric scenarios~\cite{Giudice:2004tc,Arkani-Hamed:2004zhs}. Long-lived colored particles would hadronize, forming so-called R-hadrons~\cite{Farrar:1978xj}. These bound states are comprised of the long-lived state and light SM quarks or gluons, and interact with the detector material, typically inside the calorimeters, via hadronic interactions of the light-quark constituents. Motivated by split-supersymmetric models, R-hadrons are extensively searched for at the LHC~\cite{ATLAS:2019gqq,CMS:2020iwv}. Non-observation of any  deviations of the signal from the expected background puts to a lower limit on the mass of the long-lived
$\Sigma_a$ fermion  of 2000 GeV~\cite{ATLAS:2019gqq}.

\begin{figure}[p]
    \centering
    \includegraphics[width=5cm]{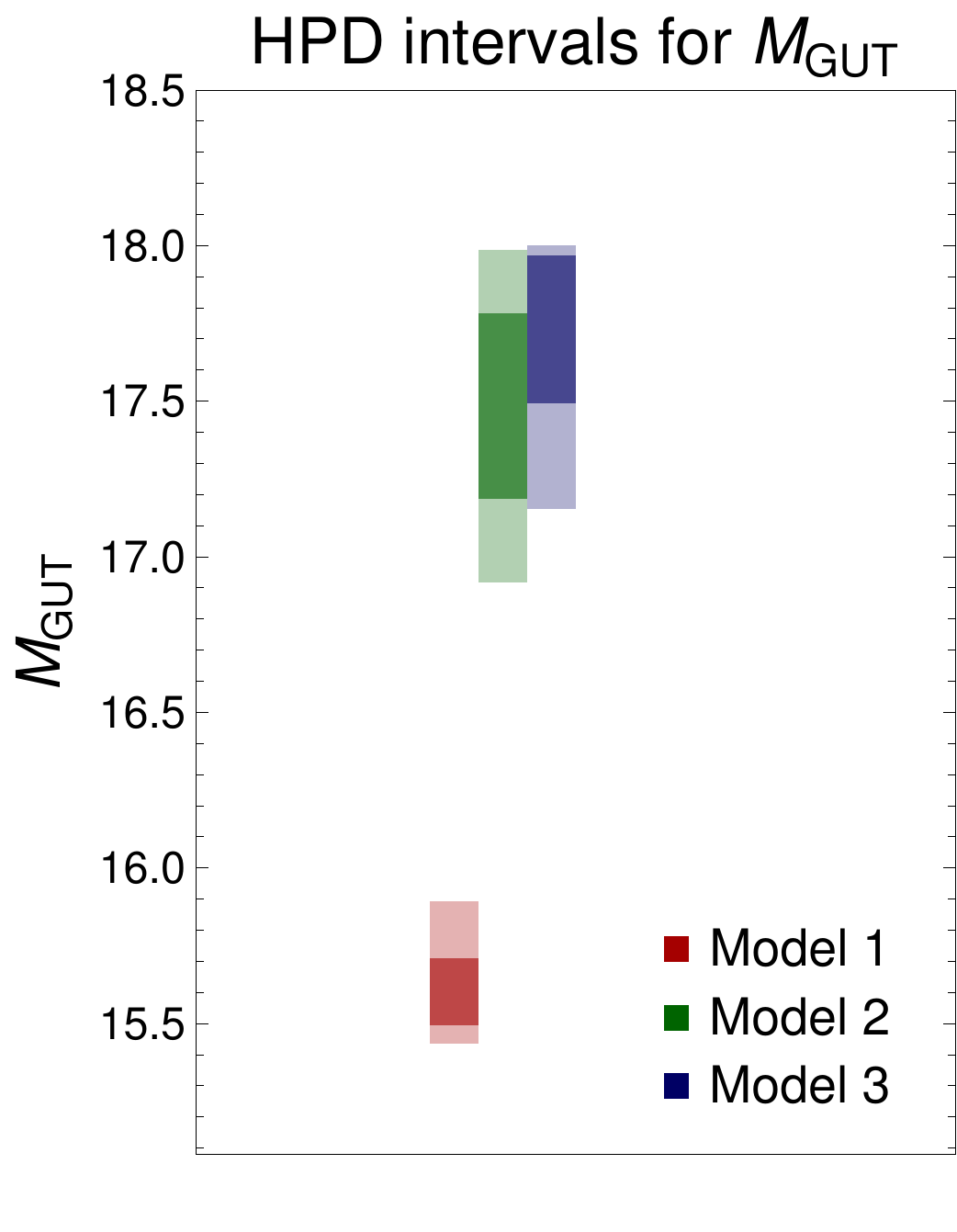}
    \caption{Predicted HPD intervals of the GUT scale. The 1$\sigma$ (2$\sigma$) HDP intervals are colored dark (light).   }
    \label{fig:MGUT}
    \vspace{10mm}
    \includegraphics[width=\textwidth]{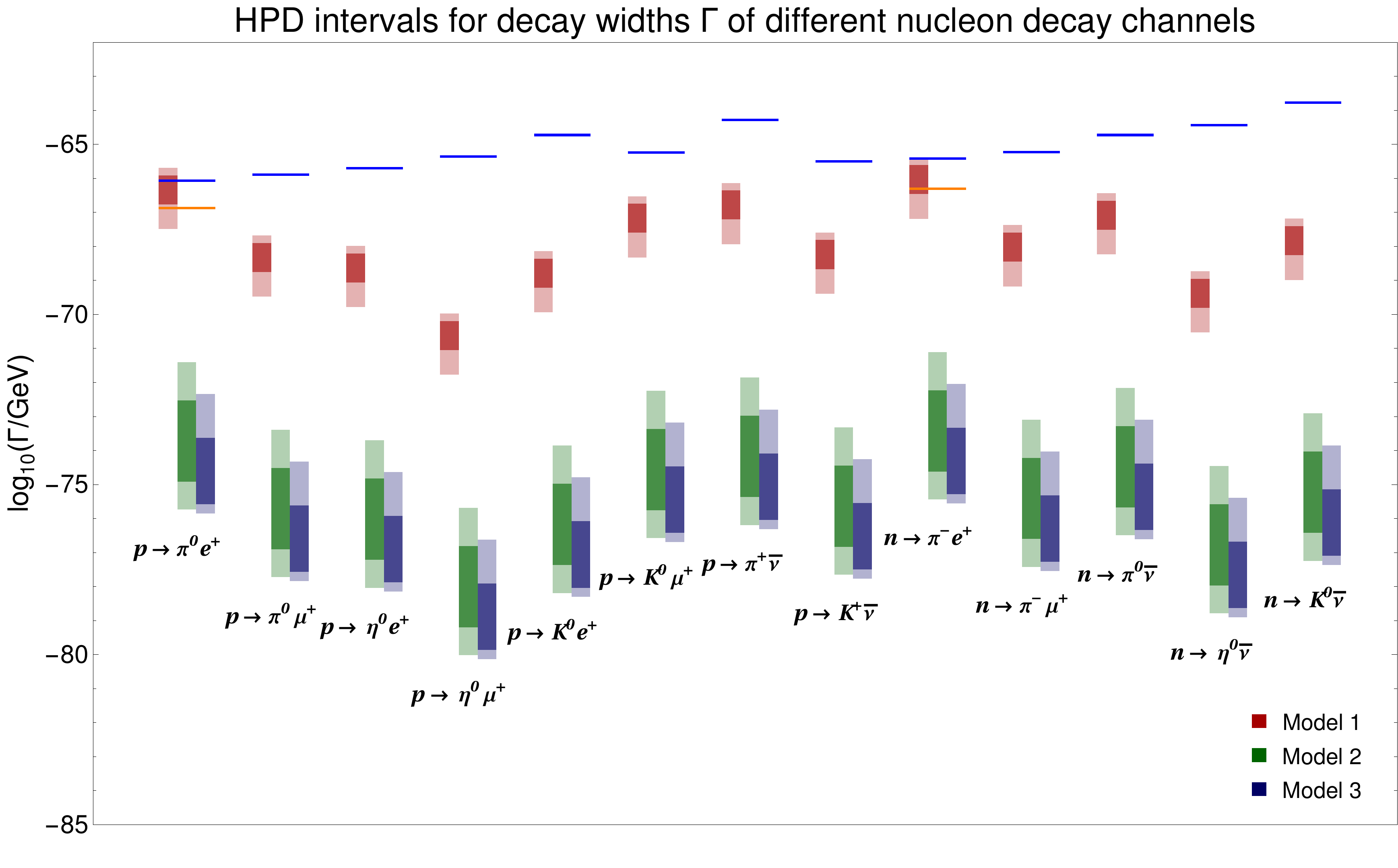}
    \caption{Predicted HPD intervals of the nucleon decay widths. The 1$\sigma$ (2$\sigma$) HDP intervals are colored dark (light). For each decay channel the blue line segments represent the current experimental constraints. The future Hyper-Kamiokande constraints for the decay channels $p\rightarrow\pi^0e^+$ and $n\rightarrow\pi^-e^+$ are indicated by orange line segments.}
    \label{fig:decaywidhts}
\end{figure}
\subsubsection{Nucleon decay width and GUT scale}
Figure\ \ref{fig:MGUT} shows the predictions for the HPD intervals of the GUT scale $M_{\text{GUT}}$. Moreover, the predicted HPD ranges for the nucleon decay widths of the various decay channels are presented in Figure\ \ref{fig:decaywidhts}. The blue line segments in the latter picture indicate the current experimental bounds at 90\ \% confidence level (cf.\ Table\ \ref{tab:nucleon decay}). Moreover, the future constraints on the decay widths for the decay channels $p\rightarrow \pi^0e^+$ and $n\rightarrow \pi^-e^+$ which will be provided by Hyper-Kamiokande\ \cite{Hyper-Kamiokande:2018ofw} are presented by orange line segments.

In Figure\ \ref{fig:MGUT} it can be seen that Model 1 clearly predicts the GUT scale to be below $10^{16}$\ GeV. On the other hand, a much larger GUT scale is preferred by the Models 2 and 3. Since the nucleon decay width is inversely proportional to the forth power of the GUT scale in the case of gauge boson mediated nucleon decay, this also results in strongly different prediction for the nucleon decay widths of the various channels as it can be seen in Figure\ \ref{fig:decaywidhts}. The nucleon decay predictions for Model 1 are very close to the current bounds, the 1$\sigma$ HPD interval of the proton decay channel $p\rightarrow \pi^0e^+$ will be fully probed by Hyper-Kamiokande. Moreover, Hyper-Kamiokande will probe most of the 1$\sigma$ HPD interval of the neutron decay channel $n\rightarrow \pi^-e^+$. On the other hand, the gauge boson mediated nucleon decay is highly suppressed in Models 2 and 3 and cannot be probed by any planed experiments. Therefore, observation of nucleon decay in the decay channels $p\rightarrow \pi^0e^+$ and $n\rightarrow \pi^-e^+$ would clearly favour Model 1 over the Models 2 and 3.

\section{Conclusion}\label{sec:conclusion}
In this paper we considered an extension of the Georgi-Glashow SU(5) GUT scenario by a 45-dimensional scalar and a 24-dimensional fermionic representation. Neutrino masses in this scenario are generated by a combination of a type I and a type III seesaw mechanism. Assuming the concept of \textit{single operator dominance} we investigated which GUT scale charged lepton and down-type quark Yukawa ratios can be viable for the second and third family and found that three combinations work: (i) $y_\tau/y_b=3/2$, $y_\mu/y_s=9/2$, (ii) $y_\tau/y_b=2$, $y_\mu/y_s=9/2$, and (iii) $y_\tau/y_b=2$, $y_\mu/y_s=6$. Also taking into account the origin of the first family masses we extended these possibilities to three toy models and analyzed various of their predictions. We showed that experimental discrimination between these models could be possible since they predict different nucleon decay rates as well as distinct light relics. 

\appendix
\appendixpage 
\section{Definition of new Yukawa couplings}\label{sec:AppendixA}
The Lagrangian density contains the two terms
\begin{align}
\mathcal{L}\supset Y_{A}\, \bm{\overline 5_F}^i \bm{24_F 5_H} 
+Y_{B}\, \bm{\overline 5_F}^i \bm{24_F 45_H}.
\end{align}
After the GUT symmetry breaking they decompose into
\begin{align}
\mathcal{L} =
\;&
\sqrt{\frac{2}{15}} Y_{A}\, d^c\Sigma_c T_a -
\sqrt{\frac{3}{10}} Y_{A}\, \ell\Sigma_c H_a +
Y_{A}\, d^c\Sigma_a  T_a+
Y_{A}\, \ell\Sigma_b  H_a+
\nonumber\\&
Y_{A}\, d^c\Sigma_d  H_a+
Y_{A}\, \ell\Sigma_e T_a +
\sqrt{\frac{5}{12}} Y_{B}\,  d^c\Sigma_c T_b+
\frac{\sqrt{5}}{4} Y_{B}\,  \ell\Sigma_c H_b+
\nonumber\\&
\frac{1}{2\sqrt{2}} Y_{B}\,  d^c\Sigma_a T_b+
\frac{1}{\sqrt{2}} Y_{B}\,  d^c\Sigma_a \phi_b+
\frac{1}{\sqrt{2}} Y_{B}\,  \ell\Sigma_a \phi_a+
\frac{1}{\sqrt{2}} Y_{B}\,  d^c\Sigma_b \phi_c+
\nonumber\\&
\frac{\sqrt{3}}{4} Y_{B}\,  \ell\Sigma_b H_b-
\frac{1}{\sqrt{2}} Y_{B}\,  d^c\Sigma_d \phi_a-
\frac{1}{2\sqrt{6}} Y_{B}\,  d^c\Sigma_d H_b+
\frac{1}{\sqrt{2}} Y_{B}\,  \ell\Sigma_d \phi_e-
\nonumber\\&
\frac{1}{\sqrt{2}} Y_{B}\,  d^c\Sigma_e \phi_d+
\frac{1}{2\sqrt{2}} Y_{B}\,  \ell\Sigma_e T_b-
\frac{1}{\sqrt{2}} Y_{B}\,  \ell\Sigma_e \phi_c\, 
\nonumber\\
 \equiv\;&
Y_{1}\, d^c\Sigma_c T_a +
Y_{2}\, \ell\Sigma_c H_a +
Y_{3}\, d^c\Sigma_a  T_a+
Y_{4}\, \ell\Sigma_b  H_a+
\nonumber\\&
Y_{5}\, d^c\Sigma_d  H_a+
Y_{6}\, \ell\Sigma_e T_a +
Y_{7}\,  d^c\Sigma_c T_b+
Y_{8}\,  \ell\Sigma_c H_b+
\nonumber\\&
Y_{9}\,  d^c\Sigma_a T_b+
Y_{10}\,  d^c\Sigma_a \phi_b+
Y_{11}\,  \ell\Sigma_a \phi_a+
Y_{12}\,  d^c\Sigma_b \phi_c+
\nonumber\\&
Y_{13}\,  \ell\Sigma_b H_b+
Y_{14}\,  d^c\Sigma_d \phi_a+
Y_{15}\,  d^c\Sigma_d H_b+
Y_{16}\,  \ell\Sigma_d \phi_e+
\nonumber\\&
Y_{17}\,  d^c\Sigma_e \phi_d+
Y_{18}\,  \ell\Sigma_e T_b+
Y_{19}\,  \ell\Sigma_e \phi_c\;, \label{decomposed}
\end{align}
where we defined the Yukawa matrices $Y_N$, with $N=1,\dots,19$.

\section{Renormalization group equations}\label{sec:AppendixB}
Here the RGEs for the gauge and Yukawa couplings as well as for the effective neutrino mass operator are listed. We have used the Mathematica package \texttt{SARAH}\ \cite{Staub:2008uz,Staub:2013tta} to obtain the RGEs for the gauge and Yukawa couplings. The SM contribution for the RGE of the effective neutrino mass operator is taken from\ \cite{Antusch:2001ck}. In order to compute the new contribution for this RGE we have used the method described therein. We use the following definition for the Heaviside-Theta function
\begin{align}
\mathcal{H}(\mu,m)=
\begin{cases}
1, \;\;\textrm{for}\;\mu\geq m,\\ 
0, \;\;\textrm{for}\;\mu < m.    
\end{cases}
\end{align}

\subsection{Gauge couplings}
The RGEs for gauge couplings ($i,k= 1-3$) are given by
\begin{align}
&\mu\frac{dg_i}{d\mu}=\frac{\beta^{g_i}_{1-\text{loop}}}{16\pi^2} +\frac{\beta^{g_i}_{2-\text{loop}}}{(16\pi^2)^2}\,,
\end{align}
 where $\beta^{g_i}_{1-\text{loop}}$ is the 1-loop and $\beta^{g_i}_{2-\text{loop}}$ is the 2-loop contribution given by
\begin{align}
&\beta^{g_i}_{1-\text{loop}}= \bigg\{ a_i^{\text{SM}}+ \mathcal{H}(\mu,m) \Delta a_i \bigg\}  g_i^3\,
\\
&\beta^{g_i}_{2-\text{loop}}=  \sum_k  b^{\text{SM}}_{ik}g^2_k + \sum_k  \Delta b_{ik}g^2_k  \;\mathcal{H}(\mu,m)+\beta^{Y,\text{SM}}_{i}+\Delta\beta^{Y}_{i}.
\end{align}
Here, $a_i^{\text{SM}}$, $b^{SM}_{ik}$ and $\beta^{Y,\text{SM}}_{i}$ are the well known SM 1-loop and 2-loop coefficients as well as Yukawa contributions\ \cite{Cheng:1973nv,Machacek:1983fi}. Moreover, the $\Delta\beta^{Y}_{i}$ are given by
\begin{align}
&\Delta\beta^{Y}_{i}= g_i^3
\sum_{k} c_{ik} Y^T_k Y^\ast_k \;\mathcal{H}^2_{k},
\end{align}
where we introduced the abbreviation $\mathcal{H}^2_{k}= \mathcal{H}(\mu,m_F) \mathcal{H}(\mu,m_H)$ associated to each of the Yukawa interactions, where, $F$ and $H$ refer to the BSM fermion and scalar appearing in that interaction, respectively, and where the $c_{ik}$ are given by
\begin{align}
&c_{1k}= -\bigg\{\frac{1}{5}, \frac{3}{10}, \frac{8}{15}, \frac{9}{20}, \frac{29}{10}, \frac{17}{5}, \frac{1}{5}, \frac{3}{10}, \frac{8}{15}, \frac{8}{15}, \frac{12}{5}, \frac{3}{5}, \frac{9}{20}, \frac{116}{15}, \frac{29}{10}, \frac{17}{5}, \frac{29}{5}, \frac{17}{5}, \frac{51}{10}   \bigg\}   ,
\\&
c_{2k}= -\bigg\{0,\frac{1}{2},0, \frac{11}{4}, \frac{3}{2}, 3,0, \frac{1}{2},0,0, 4, 6, \frac{11}{4}, 4, \frac{3}{2}, 3, 3, 3, \frac{9}{2}   \bigg\}   ,
\\&
c_{3k}= -\bigg\{\frac{1}{2},0, \frac{13}{3},0, 2, 1, \frac{1}{2},0, \frac{13}{3}, \frac{13}{3}, 6, \frac{3}{2},0, \frac{16}{3}, 2, 1, 4, 1, \frac{3}{2}   \bigg\} .  
\end{align}
Finally, the $\Delta a_i$ and $\Delta b_i$ are given as a sum over the 1-loop and 2-loop coefficients of the BSM fermions and scalars, i.e.
\begin{align}
\Delta a_i=\sum_I \Delta a_i^I\,,\qquad \Delta b_i=\sum_I \Delta b_i^I\,,
\end{align}
where $I$ runs over all BSM particles. The 1-loop coefficients are then given by
\begin{align}
    &\Delta a_i^{\phi_a}=\left\{\frac{4}{5},\frac{4}{3},2\right\},\quad
\Delta a_i^{\phi_b}=\left\{\frac{2}{15},0,\frac{5}{6}\right\},\quad
\Delta a_i^{\phi_c}=\left\{\frac{1}{5},2,\frac{1}{2}\right\},\nonumber\\&
\Delta a_i^{\phi_d}=\left\{\frac{49}{30},\frac{1}{2},\frac{1}{3}\right\},\quad
\Delta a_i^{\phi_e}=\left\{\frac{16}{15},0,\frac{1}{6}\right\},\quad
\Delta a_i^{\Phi_a}=\{0,0,\frac{1}{2}\},\nonumber\\&
\Delta a_i^{\Phi_b}=\left\{0,\frac{1}{3},0\right\},\quad
\Delta a_i^{\Sigma_a}=\{0,0,2\},\quad
\Delta a_i^{\Sigma_b}=\left\{0,\frac{4}{3},0\right\},\nonumber\\&
\Delta a_i^{\Sigma_{d,e}}=\left\{\frac{5}{3},1,\frac{2}{3}\right\},\quad
\Delta a_i^{h^\perp}=\left\{\frac{1}{10},\frac{1}{6},0\right\},\quad
\Delta a_i^{t,t^\perp}=\left\{\frac{1}{15},0,\frac{1}{6}\right\},
\end{align}
whereas the 2-loop coefficients read
\begin{align}
    &\Delta b_{ik}^{\phi_a}=
\left(
\begin{array}{ccc}
 \frac{36}{25} & \frac{36}{5} & \frac{144}{5} \\
 \frac{12}{5} & \frac{52}{3} & 48 \\
 \frac{18}{5} & 18 & 84 \\
\end{array}
\right),\quad
\Delta b_{ik}^{\phi_b}=
\left(
\begin{array}{ccc}
 \frac{8}{75} & 0 & \frac{16}{3} \\
 0 & 0 & 0 \\
 \frac{2}{3} & 0 & \frac{115}{3} \\
\end{array}
\right),\quad
\Delta b_{ik}^{\phi_c}=
\left(
\begin{array}{ccc}
 \frac{4}{25} & \frac{24}{5} & \frac{16}{5} \\
 \frac{8}{5} & 56 & 32 \\
 \frac{2}{5} & 12 & 11 \\
\end{array}
\right),\nonumber\\&
\Delta b_{ik}^{\phi_d}=
\left(
\begin{array}{ccc}
 \frac{2401}{150} & \frac{147}{10} & \frac{392}{15} \\
 \frac{49}{10} & \frac{13}{2} & 8 \\
 \frac{49}{15} & 3 & \frac{22}{3} \\
\end{array}
\right),\quad
\Delta b_{ik}^{\phi_e}=
\left(
\begin{array}{ccc}
 \frac{1024}{75} & 0 & \frac{256}{15} \\
 0 & 0 & 0 \\
 \frac{32}{15} & 0 & \frac{11}{3} \\
\end{array}
\right),\quad
\Delta b_{ik}^{\Phi_a}=
\left(
\begin{array}{ccc}
 0 & 0 & 0 \\
 0 & 0 & 0 \\
 0 & 0 & 21 \\
\end{array}
\right),\nonumber\\&
\Delta b_{ik}^{\Phi_b}=
\left(
\begin{array}{ccc}
 0 & 0 & 0 \\
 0 & \frac{28}{3} & 0 \\
 0 & 0 & 0 \\
\end{array}
\right),\quad
\Delta b_{ik}^{\Sigma_a}=
\left(
\begin{array}{ccc}
 0 & 0 & 0 \\
 0 & 0 & 0 \\
 0 & 0 & 48 \\
\end{array}
\right),\quad
\Delta b_{ik}^{\Sigma_b}=
\left(
\begin{array}{ccc}
 0 & 0 & 0 \\
 0 & \frac{64}{3} & 0 \\
 0 & 0 & 0 \\
\end{array}
\right),\nonumber\\&
\Delta b_{ik}^{\Sigma_{d,e}}=\left(
\begin{array}{ccc}
 \frac{25}{12} & \frac{15}{4} & \frac{20}{3} \\
 \frac{5}{4} & \frac{49}{4} & 4 \\
 \frac{5}{6} & \frac{3}{2} & \frac{38}{3} \\
\end{array}
\right),\quad
\Delta b_{ik}^{h^\perp}=
\left(
\begin{array}{ccc}
 \frac{9}{50} & \frac{9}{10} & 0 \\
 \frac{3}{10} & \frac{13}{6} & 0 \\
 0 & 0 & 0 \\
\end{array}
\right),\quad
\Delta b_{ik}^{t,t^\perp}=
\left(
\begin{array}{ccc}
 \frac{4}{75} & 0 & \frac{16}{15} \\
 0 & 0 & 0 \\
 \frac{2}{15} & 0 & \frac{11}{3} \\
\end{array}
\right).
\end{align}

\subsection{Yukawa matrices}
The RGEs of the Yukawa matrices read
\begin{align}
&\mu\frac{dY_f}{d\mu}= \frac{\beta_f}{16\pi^2},
\end{align}
where $f=\{u,d,e,k\}$ and $(k=1,\dots,19)$. For the SM Yukawa matrices $Y_u$, $Y_d$  and $Y_e$ (i.e. $f={u,d,e}$) the beta functions are given by
\begin{align}
&\beta_f=\beta_f^\text{SM}+\delta \beta_f,
\end{align}
where $\beta_f^\text{SM}$ is the SM beta function\ \cite{Cheng:1973nv,Machacek:1983fi}, and where
\begin{align}
&\delta \beta_f=Y_f T_1 
+\sum_{k} a^f_{k} (Y_{k})_j (Y^T_d Y^\ast_{k})_i \;\mathcal{H}^2_{k}\,.
\end{align}
Here, we have defined $T_1$ as
\begin{align}
&T_1=Y^T_2Y^\ast_2 \;\mathcal{H}^2_2 +\frac{3}{2}Y^T_4Y^\ast_4 \;\mathcal{H}^2_4 +3Y^T_5Y^\ast_5 \;\mathcal{H}^2_5.
\end{align}
while the $a_k^f$ are given by
\begin{align}
&a_k^u=\Big\{0,0,0,0,0,0,0,0,0,0,0,0,0,0,0,0,0,0,0\Big\},\\
&a_{k}^d=\left\{\frac{1}{2},0, \frac{4}{3},0, 3,0, \frac{1}{2},0, \frac{4}{3}, \frac{4}{3},0, \frac{3}{2},0, \frac{8}{3}, 1,0, 2,0,0 \right\},\\
&a_{k}^e=\left\{0,-\frac{3}{2},0, \frac{15}{4},0,\frac{3}{2},0,\frac{1}{2},0,0,4,0,\frac{3}{4},0,0,\frac{3}{2},0,\frac{3}{2},\frac{9}{4}  \right\}.
\end{align}
In order to simplify the notation, from hereon, associated to each Yukawa $Y_i\to Y_i \;\mathcal{H}^2_i$ must be understood.
The beta function of the Yukawa matrices $Y_1,\dots, Y_{19}$ then read
\begin{align}
\beta_{1}=&\;Y_1\left( -\frac{1}{5}g^2_1-4g^2_3 +\sum_{k}a_{k}^1 Y^T_{k}Y^\ast_{k} \right) +\left(Y_dY^\dagger_d \right)Y_1
+\sum_{w}b_{w}^1 \left(Y^T_1 Y^\ast_{w}\right) Y_{w}
\nonumber \\& 
+\frac{8}{3} \left(Y^T_{3} Y^\ast_{9}\right) Y_{7}+2\left(Y^T_{6} Y^\ast_{18}\right) Y_{7},
\\
\beta_{2}=&\;Y_2\left( -\frac{9}{20}g^2_1-\frac{9}{4} g^2_2 +\sum_{k}a_{k}^2 Y^T_{k}Y^\ast_{k} +T \right) + \left(-\frac{3}{2}Y_e^T Y^\ast_e \right)Y_2
+\sum_{w}b_{w}^2 \left(Y^T_2 Y^\ast_{w}\right) Y_{w}
\nonumber \\& 
+\frac{3}{2} \left(Y^T_{4} Y^\ast_{13}\right) Y_{8}+3\left(Y^T_{5} Y^\ast_{15}\right) Y_{8},
\\
\beta_{3}=&\;Y_3\left( -\frac{1}{5}g^2_1-13 g^2_3 +\sum_{k}a_{k}^3 Y^T_{k}Y^\ast_{k} \right) + \left(Y_d Y^\dagger_d \right)Y_3
+\sum_{w}b_{w}^3 \left(Y^T_3 Y^\ast_{w}\right) Y_{w}
\nonumber \\& 
+\left(Y^T_{1} Y^\ast_{7}\right) Y_{9}+2\left(Y^T_{6} Y^\ast_{18}\right) Y_{9},
\\
\beta_{4}=&\;Y_4\left( -\frac{9}{20}g^2_1-\frac{33}{4} g^2_2 +\sum_{k}a_{k}^4 Y^T_{k}Y^\ast_{k}+T \right) + \left(\frac{5}{2} Y_e^T Y^\ast_e \right)Y_4
+\sum_{w}b_{w}^4 \left(Y^T_4 Y^\ast_{w}\right) Y_{w}
\nonumber \\& 
+\left(Y^T_{2} Y^\ast_{8}\right) Y_{13}+3 \left(Y^T_{5} Y^\ast_{15}\right) Y_{13},
\\
\beta_{5}=&\;Y_5\left( -\frac{29}{20}g^2_1-\frac{9}{4} g^2_2-8g^2_3 +\sum_{k}a_{k}^5 Y^T_{k}Y^\ast_{k}+T \right) + \left(3 Y_d Y^\dagger_d \right)Y_5
+\sum_{w}b_{w}^5 \left(Y^T_5 Y^\ast_{w}\right) Y_{w}
\nonumber \\& 
+\left(Y^T_{2} Y^\ast_{8}\right) Y_{15}+\frac{3}{2} \left(Y^T_{4} Y^\ast_{13}\right) Y_{15},
\\
\beta_{6}=&\;Y_6\left( -\frac{17}{10}g^2_1-\frac{9}{2} g^2_2-4g^2_3 +\sum_{k}a_{k}^6 Y^T_{k}Y^\ast_{k} \right) + \left(\frac{1}{2} Y^T_e Y^\ast_e \right)Y_6
+\sum_{w}b_{w}^6 \left(Y^T_6 Y^\ast_{w}\right) Y_{w}
\nonumber \\& 
+\left(Y^T_{1} Y^\ast_{7}\right) Y_{18}+\frac{8}{3} \left(Y^T_{3} Y^\ast_{9}\right) Y_{18},
\\
\beta_{7}=&\;Y_7\left( -\frac{1}{5}g^2_1-4g^2_3 +\sum_{k}a_{k}^7 Y^T_{k}Y^\ast_{k} \right) + \left( Y_d Y^\dagger_d \right)Y_7
+\sum_{w}b_{w}^7 \left(Y^T_7 Y^\ast_{w}\right) Y_{w}
\nonumber \\& 
+\left(2Y^T_{18} Y^\ast_{6}\right) Y_{1}+\frac{8}{3} \left(Y^T_{9} Y^\ast_{3}\right) Y_{1},
\\
\beta_{8}=&\;Y_8\left( -\frac{9}{20}g^2_1-\frac{9}{4} g^2_2 +\sum_{k}a_{k}^8 Y^T_{k}Y^\ast_{k} \right) + \left( Y_e^T Y^\ast_e \right)Y_8
+\sum_{w}b_{w}^8 \left(Y^T_8 Y^\ast_{w}\right) Y_{w}
\nonumber \\& 
+\frac{3}{2} \left(Y^T_{13} Y^\ast_{4}\right) Y_{2}+3 \left(Y^T_{15} Y^\ast_{5}\right) Y_{2},
\\
\beta_{9}=&\;Y_9\left( -\frac{1}{5}g^2_1-13g^2_3 +\sum_{k}a_{k}^9 Y^T_{k}Y^\ast_{k} \right) + \left( Y_d Y^\dagger_d \right)Y_9
+\sum_{w}b_{w}^9 \left(Y^T_9 Y^\ast_{w}\right) Y_{w}
\nonumber \\& 
+2\left(Y^T_{18} Y^\ast_{6}\right) Y_{3}+ \left(Y^T_{7} Y^\ast_{1}\right) Y_{3},
\\
\beta_{10}=&\;Y_{10}\left( -\frac{1}{5}g^2_1-13g^2_3 +\sum_{k}a_{k}^{10} Y^T_{k_{10}}Y^\ast_{k} \right) + \left( Y_d Y^\dagger_d \right)Y_{10}
+\sum_{w}b_{w}^{10} \left(Y^T_{10} Y^\ast_{w}\right) Y_{w},
\\
\beta_{11}=&\;Y_{11}\left( -\frac{9}{20}g^2_1-\frac{9}{4}g^2_2-9g^2_3 +\sum_{k}a_{k}^{11} Y^T_{k}Y^\ast_{k} \right) + \left( Y_e^T Y^\ast_e \right)Y_{11}
+\sum_{w}b_{w}^{11} \left(Y^T_{11} Y^\ast_{w}\right) Y_{w},
\\
\beta_{12}=&\;Y_{12}\left( -\frac{1}{5}g^2_1-6g^2_2-4g^2_3 +\sum_{k}a_{k}^{12} Y^T_{k}Y^\ast_{k} \right) + \left( Y_d Y^\dagger_d \right)Y_{12}
+\sum_{w}b_{w}^{12} \left(Y^T_{12} Y^\ast_{w}\right) Y_{w},
\\
\beta_{13}=&\;Y_{13}\left( -\frac{9}{20}g^2_1-\frac{33}{4} g^2_2 +\sum_{k}a_{k}^{13} Y^T_{k}Y^\ast_{k} \right) + \left(\frac{1}{2} Y_e^T Y^\ast_e \right)Y_{13}
+\sum_{w}b_{w}^{13} \left(Y^T_{13} Y^\ast_{w}\right) Y_{w}
\nonumber \\& 
+3\left(Y^T_{15} Y^\ast_{5}\right) Y_{4}+ \left(Y^T_{8} Y^\ast_{2}\right) Y_{4},
\\
\beta_{14}=&\;Y_{14}\left( -\frac{29}{20}g^2_1-\frac{9}{4}g^2_2-8g^2_3 +\sum_{k}a_{k}^{14} Y^T_{k}Y^\ast_{k} \right) + \left( Y_d Y^\dagger_d \right)Y_{14}
+\sum_{w}b_{w}^{14} \left(Y^T_{14} Y^\ast_{w}\right) Y_{w},
\\
\beta_{15}=&\;Y_{15}\left( -\frac{29}{20}g^2_1-\frac{9}{4}g^2_2-8g^2_3 +\sum_{k}a_{k}^{15} Y^T_{k}Y^\ast_{k} \right) + \left( Y_d Y^\dagger_d \right)Y_{15}
+\sum_{w}b_{w}^{15} \left(Y^T_{15} Y^\ast_{w}\right) Y_{w}
\nonumber \\& 
+\frac{3}{2} \left(Y^T_{13} Y^\ast_{4}\right) Y_{5}+ \left(Y^T_{8} Y^\ast_{2}\right) Y_{5},
\\
\beta_{16}=&\;Y_{16}\left( -\frac{17}{10}g^2_1 -\frac{9}{2}g^2_2-4g^2_3 +\sum_{k}a_{k}^{16} Y^T_{k}Y^\ast_{k} \right) + \left(\frac{1}{2} Y_e^T Y^\ast_e \right)Y_{16}
+\sum_{w}b_{w}^{16} \left(Y^T_{16} Y^\ast_{w}\right) Y_{w},
\\
\beta_{17}=&\;Y_{17}\left( -\frac{29}{20}g^2_1-\frac{9}{4}g^2_2-8g^2_3 +\sum_{k}a_{k}^{17} Y^T_{k}Y^\ast_{k} \right) + \left( Y_d Y^\dagger_d \right)Y_{17}
+\sum_{w}b_{w}^{17} \left(Y^T_{17} Y^\ast_{w}\right) Y_{w},
\\
\beta_{18}=&\;Y_{18}\left( -\frac{17}{10}g^2_1  -\frac{9}{2}g^2_2 -4g^2_3 +\sum_{k}a_{k}^{18} Y^T_{k}Y^\ast_{k} \right) + \left(\frac{1}{2} Y_e^T Y^\ast_e \right)Y_{18}
+\sum_{w}b_{w}^{18} \left(Y^T_{18} Y^\ast_{w}\right) Y_{w}
\nonumber \\& 
+\left(Y^T_{7} Y^\ast_{1}\right) Y_{6}+\frac{8}{3} \left(Y^T_{9} Y^\ast_{3}\right) Y_{6},
\\
\beta_{19}=&\;Y_{19}\left( -\frac{17}{20}g^2_1 -\frac{9}{2}g^2_2-4g^2_3 +\sum_{k}a_{k}^{19} Y^T_{k}Y^\ast_{k} \right) + \left(\frac{1}{2} Y_e^T Y^\ast_e \right)Y_{19}
+\sum_{w}b_{w}^{19} \left(Y^T_{19} Y^\ast_{w}\right) Y_{w},
\end{align}
where the coefficients $a_k^f$ are given by
\begin{align}
&a_{k}^1=\{3,1,\frac{8}{3},0,0,2,\frac{3}{2},1,0,0,0,0,0,0,0,0,0,0,0\},
\\
&a_{k}^2=\{\frac{3}{2},\frac{5}{2},0,\frac{3}{2},3,0,\frac{3}{2},1,0,0,0,0,0,0,0,0,0,0,0\},
\\
&a_{k}^3=\{1,0,\frac{9}{2},0,0,2,0,0,\frac{1}{2},\frac{1}{2},1,0,0,0,0,0,0,0,0\},
\\
&a_{k}^4=\{0,1,0,\frac{11}{4},3,0,0,0,0,0,0,\frac{3}{2},\frac{1}{2},0,0,0,0,0,0\},
\\
&a_{k}^5=\{0,1,0,\frac{3}{2},\frac{9}{2},0,0,0,0,0,0,0,0,\frac{4}{3},\frac{1}{2},\frac{1}{2},0,0,0\},
\\
&a_{k}^6=\{1,0,\frac{8}{3},0,0,4,0,0,0,0,0,0,0,0,0,0,1,\frac{1}{2},\frac{3}{4}\},
\\
&a_{k}^7=\{\frac{3}{2},1,0,0,0,0,3,1,\frac{8}{3},0,0,0,0,0,0,0,0,2,0\},
\\
&a_{k}^8=\{\frac{3}{2},1,0,0,0,0,\frac{3}{2},\frac{5}{2},0,0,0,0,\frac{3}{2},0,3,0,0,0,0\},
\\
&a_{k}^9=\{0,0,\frac{1}{2},0,0,0,1,0,\frac{9}{2},\frac{1}{2},1,0,0,0,0,0,0,2,0\},
\\
&a_{k}^{10}=\{0,0,\frac{1}{2},0,0,0,0,0,\frac{1}{2},\frac{19}{6},1,0,0,0,0,0,0,0,0\},
\\
&a_{k}^{11}=\{0,0,\frac{1}{2},0,0,0,0,0,\frac{1}{2},\frac{1}{2},6,0,0,1,0,0,0,0,0\},
\\
&a_{k}^{12}=\{0,0,0,\frac{1}{2},0,0,0,0,0,0,0,4,\frac{1}{2},0,0,0,0,0,1\},
\\
&a_{k}^{13}=\{0,0,0,\frac{1}{2},0,0,0,1,0,0,0,\frac{3}{2},\frac{11}{4},0,3,0,0,0,0\},
\\
&a_{k}^{14}=\{0,0,0,0,\frac{1}{2},0,0,0,0,0,1,0,0,5,\frac{1}{2},\frac{1}{2},0,0,0\},
\\
&a_{k}^{15}=\{0,0,0,0,\frac{1}{2},0,0,1,0,0,0,0,\frac{3}{2},\frac{4}{3},\frac{9}{2},\frac{1}{2},0,0,0\},
\\
&a_{k}^{16}=\{0,0,0,0,\frac{1}{2},0,0,0,0,0,0,0,0,\frac{4}{3},\frac{1}{2},4,0,0,0\},
\\
&a_{k}^{17}=\{0,0,0,0,0,\frac{1}{2},0,0,0,0,0,0,0,0,0,0,5,\frac{1}{2},\frac{3}{4}\},
\\
&a_{k}^{18}=\{0,0,0,0,0,\frac{1}{2},1,0,\frac{8}{3},0,0,0,0,0,0,0,1,4,\frac{3}{4}\},
\\
&a_{k}^{19}=\{0,0,0,0,0,\frac{1}{2},0,0,0,0,0,1,0,0,0,0,1,\frac{1}{2},4\},
\end{align}
while the coefficients $b_{k}^f$ read
\begin{align}
&b_{w}^1=\{0,0,\frac{4}{3},0,1,0,\frac{3}{2},0,0,\frac{4}{3},0,\frac{3}{2},0,\frac{8}{3},1,0,2,0,0\},\\
&b_{w}^2=\{0,0,0,\frac{3}{4},0,\frac{3}{2},0,\frac{3}{2},0,0,4,0,\frac{3}{4},0,0,\frac{3}{2},0,\frac{3}{2},\frac{9}{4}\},\\
&b_{w}^3=\{\frac{1}{2},0,0,0,1,0,\frac{1}{2},0,0,\frac{4}{3},0,\frac{3}{2},0,\frac{8}{3},1,0,2,0\},\\
&b_{w}^4=\{0,\frac{1}{2},0,0,0,\frac{3}{2},0,\frac{1}{2},0,0,4,0,\frac{9}{4},0,0,\frac{3}{2},0,\frac{3}{2},\frac{9}{4}\},\\
&b_{w}^5=\{\frac{1}{2},0,\frac{4}{3},0,0,0,\frac{1}{2},0,\frac{4}{3},\frac{4}{3},0,\frac{3}{2},0,\frac{8}{3},4,0,2,0,0\},\\
&b_{w}^6=\{0,\frac{1}{2},0,\frac{3}{4},0,0,0,\frac{1}{2},0,0,4,0,\frac{3}{4},0,0,\frac{3}{2},0,\frac{7}{2},\frac{9}{4}\},\\
&b_{w}^7=\{\frac{3}{2},0,\frac{4}{3},0,1,0,0,0,\frac{4}{3},\frac{4}{3},0,\frac{3}{2},0,\frac{8}{3},1,0,2,0,0\},\\
&b_{w}^8=\{0,\frac{3}{2},0,\frac{3}{4},0,\frac{3}{2},0,0,0,0,4,0,\frac{3}{4},0,0,\frac{3}{2},0,\frac{3}{2},\frac{9}{4}\},\\
&b_{w}^9=\{\frac{1}{2},0,4,0,1,0,\frac{1}{2},0,0,\frac{4}{3},0,\frac{3}{2},0,\frac{8}{3},1,0,2,0,0\},\\
&b_{w}^{10}=\{\frac{1}{2},0,\frac{4}{3},0,1,0,\frac{1}{2},0,\frac{4}{3},0,0,\frac{3}{2},0,\frac{8}{3},1,0,2,0,0\},\\
&b_{w}^{11}=\{0,\frac{1}{2},0,\frac{3}{4},0,\frac{3}{2},0,\frac{1}{2},0,0,0,0,\frac{3}{4},0,0,\frac{3}{2},0,\frac{3}{2},\frac{9}{4}\},\\
&b_{w}^{12}=\{\frac{1}{2},0,\frac{4}{3},0,1,0,\frac{1}{2},0,\frac{4}{3},\frac{4}{3},0,0,0,\frac{8}{3},1,0,2,0,0\},\\
&b_{w}^{13}=\{0,\frac{1}{2},0,\frac{9}{4},0,\frac{3}{2},0,\frac{1}{2},0,0,4,0,0,0,0,\frac{3}{2},0,\frac{3}{2},\frac{9}{4}\},\\
&b_{w}^{14}=\{\frac{1}{2},0,\frac{4}{3},0,1,0,\frac{1}{2},0,\frac{4}{3},\frac{4}{3},0,\frac{3}{2},0,0,1,0,2,0,0\},\\
&b_{w}^{15}=\{\frac{1}{2},0,\frac{4}{3},0,4,0,\frac{1}{2},0,\frac{4}{3},\frac{4}{3},0,\frac{3}{2},0,\frac{8}{3},0,0,2,0,0\},\\
&b_{w}^{16}=\{0,\frac{1}{2},0,\frac{3}{4},0,\frac{3}{2},0,\frac{1}{2},0,0,4,0,\frac{3}{4},0,0,0,0,\frac{3}{2},\frac{9}{4}\},\\
&b_{w}^{17}=\{\frac{1}{2},0,\frac{4}{3},0,1,0,\frac{1}{2},0,\frac{4}{3},\frac{4}{3},0,\frac{3}{2},0,\frac{8}{3},1,0,0,0,0\},\\
&b_{w}^{18}=\{0,\frac{1}{2},0,\frac{3}{4},0,\frac{7}{2},0,\frac{1}{2},0,0,4,0,\frac{3}{4},0,0,\frac{3}{2},0,0,\frac{9}{4}\},\\
&b_{w}^{19}=\{0,\frac{1}{2},0,\frac{3}{4},0,\frac{3}{2},0,\frac{1}{2},0,0,4,0,\frac{3}{4},0,0,\frac{3}{2},0,\frac{3}{2},0\}.
\end{align}

\subsection{Effective neutrino mass operator}
The RGE for the effective neutrino mass operator reads
\begin{align}
    16\pi^2 \mu\frac{d\kappa}{d\mu}=&\;\beta_\kappa^{\text{SM}}+\Delta\beta_\kappa,
\end{align}
where $\beta_\kappa^{\text{SM}}$ is the SM contribution as given in\ \cite{Antusch:2001ck} and $\Delta\beta_\kappa$ is the correction due to the added BSM particles. For $\Delta\beta_\kappa$ we find\footnote{To simplify the analysis we ignore RG induced mixings between different dimension five operators.}
\begin{align}
    \Delta\beta_\kappa=\kappa\bigg(&\frac{1}{2}Y_2^* Y_2^T + \frac{3}{4}Y_4^*Y_4^T+ \frac{3}{2}Y_6^*Y_6^T+ \frac{1}{2}Y_8^*Y_8^T+ 4Y_{11}^*Y_{11}^T+ \frac{3}{4}Y_{13}^*Y_{13}^T+ \frac{3}{2}Y_{16}^*Y_{16}^T \nonumber\\ 
    +&\; \frac{3}{2}Y_{18}^*Y_{18}^T+ \frac{9}{4}Y_{19}^*Y_{19}^T\bigg)+ \bigg(\frac{1}{2}Y_2^* Y_2^T + \frac{3}{4}Y_4^*Y_4^T+ \frac{3}{2}Y_6^*Y_6^T + \frac{1}{2}Y_8^*Y_8^T \nonumber\\ 
    +&\; 4Y_{11}^*Y_{11}^T + \frac{3}{4}Y_{13}^*Y_{13}^T+ \frac{3}{2}Y_{16}^*Y_{16}^T+ \frac{3}{2}Y_{18}^*Y_{18}^T+ \frac{9}{4}Y_{19}^*Y_{19}^T\bigg)^T \kappa \nonumber\\
    +&\;\bigg(2Y_2^TY_2^*+ 3Y_4^TY_4^*+ 6Y_5^TY_5^*+ 2Y_8^TY_8^*+ 3Y_{13}^TY_{13}^*+ 6Y_{15}^TY_{15}^*\bigg)\kappa \;.
\end{align}


\bibliographystyle{style}
\bibliography{references}

\end{document}